\documentclass[11pt]{article}
\usepackage{graphicx, amsmath, amssymb, physics, caption, subcaption} % Required for inserting images
\usepackage[dvipsnames]{xcolor}
\usepackage{amsfonts}
\usepackage{amssymb}
\usepackage{mathrsfs}
\usepackage{graphicx}
\usepackage{amsmath,amsthm,amssymb,amscd}
\usepackage[all]{xy}
\usepackage{multirow}
\usepackage{color}
\usepackage{float}
\usepackage{mathtools,slashed,tensor}
\usepackage{bbold}
\usepackage{bbm}
\usepackage{slashed}
\usepackage{cancel}
\usepackage[normalem]{ulem}
\usepackage{cite, comment}

\usepackage{jheppub}

\def\be{\begin{equation}}
\def\ee{\end{equation}}
\def\bea{\begin{eqnarray}}
\def\eea{\end{eqnarray}}

\preprint{LCTP-24-01}

\title{Universality in Logarithmic Temperature Corrections to Near-Extremal Rotating Black Hole Thermodynamics in Various Dimensions}

\author[a]{Sabyasachi Maulik,}

\author[b,c]{Leopoldo A.~Pando Zayas,}

\author[d]{Augniva Ray,}

\author[b]{Jingchao Zhang}

\emailAdd{mauliks@iitk.ac.in, lpandoz@umich.edu, augniva.ray@apctp.org, jingchaz@umich.edu }

\affiliation[a]{Department of Physics, Indian Institute of Technology Kanpur, Kalyanpur, Kanpur, Uttar Pradesh 208016, India}

\affiliation[b]{Leinweber Center for Theoretical Physics, 
University of Michigan, Ann Arbor, MI 48109, USA}

\affiliation[c]{The Abdus Salam International Centre for Theoretical Physics, 34014 Trieste, Italy}

\affiliation[d]{Asia Pacific Center for Theoretical Physics, Postech, Pohang 37673, Korea}

\abstract{The low-temperature thermodynamics of near-extremal rotating black holes has recently been revisited to incorporate  one-loop contributions that are dominant in this regime. We discuss these quantum corrections to the gravitational path integral for asymptotically Anti de-Sitter black holes in four and five dimensions. In four dimensions we explicitly consider Kerr-AdS$_4$, Kerr-Newman-AdS$_4$ and the rotating black hole in ${\cal N}=4$ gauged supergravity with two scalars and two electric charges turned on. In five  dimensions we explicitly address the asymptotically flat Myers-Perry black hole and the  Kerr-AdS$_5$ black hole. In every case we find that tensor modes contribute $\frac{3}{2} \log T_{\rm Hawking}$ to the low-temperature thermodynamics. We identify the root cause of this universality in two facts: (i) the universal presence of a $SL(2,\mathbb{R})$ subgroup of isometries in the near-horizon geometry and (ii) a set of cancellations in the Lichnerowicz operator. 
We show that these two conditions hold for near-extremal black holes in asymptotically flat and asymptotically AdS spacetimes of various dimensions.}

\begin{document}

\maketitle

\section{Introduction}

The low temperature breakdown of black hole thermodynamics for extremal black holes was pointed out more than thirty years ago \cite{Preskill:1991tb}. The conundrum consists in that, at very low temperature, the emission of one Hawking quantum can drastically change the temperature of the near-extremal black hole. It is quite remarkable that the resolution to this puzzle did not require knowledge of the full path integral (quantum gravity) and was first achieved by a careful treatment of certain zero modes in the extremal solution \cite{Iliesiu:2020qvm,Heydeman:2020hhw,Boruch:2022tno}. The realization that temperature effectively acts as a coupling constant whereby the high-temperature regime is classical while the very low-temperature regime is quantum and strongly coupled was understood first in the context of two-dimensional Jackiw-Teitelboim (JT) gravity \cite{Almheiri:2014cka,Jensen:2016pah,Maldacena:2016upp}.

The key realization  in \cite{Iliesiu:2020qvm,Heydeman:2020hhw,Boruch:2022tno} resides in understanding that the near-horizon region of the higher-dimensional black hole contains a JT subsector that dominates the full path integral. In this sense, a once interesting toy model of two-dimensional gravity finds impactful implications for the physics of higher-dimensional black holes.  
These works emphasized the role of zero modes in the full geometry and their connection with the Schwarzian action (see \cite{Mertens:2022irh,Turiaci:2023wrh} for reviews), thus pointing to universality within high-dimensional extremal black holes. Any higher-dimensional gravity theory admitting near-extremal solutions will admit such zero modes and, consequently, the low-temperature thermodynamics will be accordingly corrected.  

In practice, embedding quantum aspects of JT gravity in spherically symmetric near-extremal black holes is almost immediate. Indeed, the relevance of JT for Reissner-Nordstrom black holes is directly suggested by the fact that  the near-horizon geometry of extremal Reissner-Nordstrom is AdS$_2\times S^p$. Another recent spherically symmetric example is \cite{Emparan:2023ypa} which relies on the properties of the modes to  resolve an old puzzle concerning the entanglement entropy of holographic quantum fields in Rindler space.

For rotating black holes, however, the relation is not that immediate and has been explicitly worked out only recently  \cite{Kapec:2023ruw,Rakic:2023vhv}. Namely, the near-horizon extremal Kerr solution admits a family of normalizable zero modes corresponding to reparametrizations of the boundary time, just as in JT gravity. The path integral over these zero modes leads to an infrared divergence in the one-loop approximation to the Euclidean partition function. This divergence can be regulated following recent suggestion of  turning on a small but finite temperature correction in the  geometry \cite{Iliesiu:2022onk, Banerjee:2023quv,Banerjee:2023gll}. The resulting finite-temperature geometry lifts the eigenvalues of the zero modes, rendering the path integral infrared finite and leading to the thermodynamic-altering correction to the near-extremal black hole: $\frac{3}{2}\log T_{\rm Hawking}$.

In this manuscript,  largely motivated by applications to the AdS/CFT correspondence, we perform a similar path integral treatment for asymptotically AdS black holes. We consider Kerr-AdS$_4$, Kerr-Newman-AdS$_4$ and a near-extremal black hole in ${\cal N}=4$ gauged supergravity that contains two scalar fields and two Maxwell fields. We then turn to five dimensions where we consider the asymptotically flat Myers-Perry five-dimensional black hole as well as the Kerr-AdS$_5$ black hole, explicitly. In every case we find that the contribution to the path integral given by the gravitional tensor modes is $\frac{3}{2}\log T_{\rm Hawking}$. Along the way, we gain understanding into this universality as a consequence of symmetries preserved near the horizon and particular properties of the kinetic (Lichnerowicz) operator for tensor modes. 

An added motivation for our work are the positive results in the framework of viewing certain black holes in the AdS/CFT correspondence through the lens of the Kerr/CFT correspondence. In  \cite{David:2020ems} aspects of the  Kerr/CFT correspondence were applied to the near-extremal black holes in AdS$_{4,5,6,7}$ and yielded a microscopic explanation for their entropy, besides the canonical AdS$_{d+1}$/CFT$_d$ one. Some sub-leading aspects, such as certain logarithmic corrections to the entropy of AdS$_5$ black holes and black strings, were also successfully matched \cite{David:2021eoq}. It is natural to expect similar and complementary quantum insights to arise by championing the JT point of view for the near-horizon geometries. We hope that this synergy of two lower-dimensional approaches will shed light on important quantum aspects of higher-dimensional black holes. 
 
 %%%%%%%%%%%%%%%%%%%%%%%%%%%%%%%%%%%%%%%%%%

The rest of the manuscript is organized as follows. In section \ref{Sec:Kerr-AdS} we review the geometry of Kerr-AdS$_4$, discuss its near-horizon extremal limit and explicitly derived the low-temperature contribution to the thermodynamics. Section \ref{Sec:KN-AdS} considers Kerr-Newman-AdS$_4$ black holes. In section \ref{Sec:Sugra} we discuss the rotating black hole with two scalars and two electric charges turned on in ${\cal N}=4$ four-dimensional gauged supergravity.  Section \ref{Sec:MyersPerryBlack hole} present details of the tensor modes computation for the asymptoticaly flat five-dimensional Myers-Perry black hole while in section \ref{sec:Kerr-AdS5} we discuss the Kerr-AdS$_5$ black hole explicitly.   We collect a number of observations and remarks regarding the universality of the logarithmic in temperature correction from tensor modes in section \ref{Sec:Universality} and conclude in section \ref{Sec:Conclusions}. We relegate some technical material and a few very long expressions to a series of appendices.

%%%%%%%%%%%%%%%%%%%%%%%%%%%%%%%%%%%%%%%%%%%%%%%%%%%%%%%%%
\section{Kerr-AdS$_4$  black hole}\label{Sec:Kerr-AdS}

%%%%%%%%%%%%%%%%%%%%%%%%%%%%%%%%%%%%%%%%%%%%%%%%%%%%%%%%%%%%%%%%%%%
%\subsection{Kerr AdS black hole in four dimensions} 

The Kerr-AdS$_4$ black hole is a solution of the equations of motion of the Einstein-Hilbert action in $\left(3+1\right)$ spacetime dimensions with a negative cosmological constant (we always work in units where the Newton's constant $G_N$ = 1.)
\begin{equation}
    I_{\mathrm{grav}} = -\frac{1}{16\pi} \int_{\mathcal{M}} d^4x \sqrt{-g} \left(R - 2\Lambda \right) - \frac{1}{8\pi} \int_{\partial\mathcal{M}} d^3y \sqrt{h}\, K.
\end{equation}

\noindent The geometry is described by the metric (\cite{Carter:1968ks,Plebanski:1976gy}, we will follow the notation of  \cite{Caldarelli:1999xj})
\begin{equation} \label{eq:metricKerrAdS4}
    ds^2 = - \frac{\Delta_r}{\Sigma^2} \left(dt - \frac{a}{\Xi} \sin^2 \theta d{\phi}\right)^2 + \frac{\Sigma^2}{\Delta_r}dr^2 + \frac{\Sigma^2}{\Delta_\theta}d\theta^2 + \frac{\Delta_\theta}{\Sigma^2}\sin^2 \theta\left( a dt- \frac{\left(r^2+a^2\right)}{\Xi}d{\phi} \right)^2 \, ,
\end{equation}
where
\begin{eqnarray}
    \Delta_r &=& \left(r^2+a^2\right)\left(a+\frac{r^2}{L^2}\right) - 2 M r \, , \quad  \Delta_\theta = 1 - \frac{a^2}{L^2} \cos^2\theta\, , \\
    \Xi &=& 1 - \frac{a^2}{L^2} \, ,  \qquad \Sigma = \sqrt{r^2 + a^2 \cos^2 \theta}\, . 
\end{eqnarray}
The cosmological constant $\Lambda$ is related to the length-scale $L$ by $\Lambda = -\frac{3}{L^2}$. The black hole is characterized by two conserved quantities, namely its energy $E$ and angular momentum $J$:
\begin{equation} \label{eq:chargeKerrAdS4}
    E = \frac{M}{\Xi^2}, \qquad J = \frac{M\,a}{\Xi^2}.
\end{equation}
The thermodynamic properties of this black hole have been extensively studied in the literature, see e.g. \cite{,Caldarelli:1999xj,Gibbons:2004ai}. The outer event horizon $r_{+}$ is the largest real, positive root of $\Delta_r = 0$. The Bekenstein-Hawking entropy is related to the area of the event horizon in the usual manner, and is given by
\begin{equation}
    S = \frac{\pi\left(r_{+}^2 + a^2\right)}{\Xi},
\end{equation}
while the surface gravity $\kappa$, and the temperature are
\begin{equation} \label{eq:tempKerrAdS4}
    T = \frac{\kappa}{2\pi} = \frac{r_{+}\left(1 + \frac{a^2}{L^2} + \frac{3 r_{+}^2}{L^2} - \frac{a^2}{r_{+}^2} \right)}{4\pi\left(r_{+}^2 + a^2\right)}.
\end{equation}
The angular velocity of the black hole relative to a frame which is non-rotating at infinity, is given by
\begin{equation}
    \Omega = \frac{a\left(1+\frac{r_{+}^2}{L^2}\right)}{r_{+}^2 + a^2}.
\end{equation}
Together, these quantitites satisfy the first law of black hole thermodynamics for the outer event horizon
\begin{equation}
    dE = TdS + \Omega dJ.
\end{equation}

\subsection{Near-extremal, near-horizon geometry}

We are interested in the near horizon geometry of the Kerr-AdS$_4$ black hole for small deviations away from extremality. For this purpose, let us rewrite the function $\Delta_r$ in the black-hole metric \eqref{eq:metricKerrAdS4} as
\begin{equation}
    \Delta_r = \frac{1}{L^2}\left(r - r_{+}\right)\left(r - r_{-}\right)\left( r^2 + \left(r_{+} + r_{-}\right)r + \frac{L^2\left(L^2 + r_{+}^2 + r_{-}^2 + r_{+}r_{-} \right)}{L^2 - r_{+}r_{-}}\right), 
\end{equation}
where $r_{\pm}$ are the two real roots of $\Delta_r = 0$, they describe the outer and inner horizon radii of the black hole. In terms of these two parameters, the mass parameter and the angular momentum are expressed as
\begin{align} \label{eq:Mseries}
    M &= \frac{\left(r_{-}+r_{+}\right) \left(L^2+r_-^2\right) \left(L^2+r_+^2\right)}{2 L^2 \left(L^2-r_{+} r_{-}\right)},\\
    J &= \frac{L \sqrt{r_{+}r_{-}} \left(r_-+r_+\right) \left(L^2+r_-^2\right) \left(L^2-r_{+}r_{-}\right) \left(L^2+r_+^2\right) \sqrt{\frac{L^2 \left(L^2+r_-^2+r_+^2+r_{+}r_{-}\right)}{L^2-r_{+}r_{-}}}}{2 \left(L^4-2 L^2 r_{+}r_{-}-r_{+}r_{-} \left(r_-^2+r_+ r_-+r_+^2\right)\right){}^2}.
\end{align}
Note that $a$, which is related to $J$ by equation \eqref{eq:chargeKerrAdS4}, is given by
\begin{equation} \label{eq:aseries}
    a = \frac{L\sqrt{r_{+}r_{-}}\left(L^2 + r_{+}^2 + r_{-}^2 + r_{+}r_{-} \right)}{L^2 - r_{+}r_{-}}\, .
\end{equation}
The merit of such parameterization is that the extremal limit can be easily defined as the limit when the two horizons coalesce into one, and the temperature becomes zero. Let us denote the radius of the event horizon at extremality as $r_0$. We find that $r_{\pm}$ have the following small temperature ($T$) expansion
\begin{eqnarray}
    r_{\pm} = r_0 \pm \frac{4 \pi  L^2 r_0^2 \left(L^2+r_0^2\right)}{L^4+6 L^2 r_0^2-3 r_0^4} T \pm \frac{4 \pi ^2 L^2 r_0^3 \left(L^4+3 r_0^4\right) \left(5 L^6+11 L^4 r_0^2-17 L^2 r_0^4+9 r_0^6\right)}{\left(L^4+6 L^2 r_0^2-3 r_0^4\right)^3} T^2 +\ldots, \nonumber \\ \label{eq:rpseriesKAdS4}
\end{eqnarray}
and from equations \eqref{eq:Mseries} and \eqref{eq:aseries}, $M$ and $a$ receive corrections which are given by
\begin{eqnarray} \label{eq:MseriesKerrAdS}
    M &&= \frac{r_0 \left(L^2+r_0^2\right)^2}{L^2 \left(L^2-r_0^2\right)}+\frac{4  \left(\pi ^2 L^4 r_0^3+2 \pi ^2 L^2 r_0^5+9 \pi ^2 r_0^7\right) \left(L^2+r_0^2\right)}{\left(L^2-r_0^2\right) \left(L^4+6 L^2 r_0^2-3 r_0^4\right)}T^2 + \ldots\, ,\\
    a &&= \frac{L r_0 }{(L-r_0) (L+r_0)} \times \label{eq:aseriesKerrAdS} \\ 
    &&\left(L^2+3r_0^2-\frac{4 \pi ^2 L^2 r_0^2  \left(L^2+3 r_0^2\right) \left(L^{10}+5 L^8 r_0^2-2 L^6 r_0^4-202 L^4 r_0^6+161 L^2 r_0^8-27 r_0^{10}\right)}{\left(L^4+6 L^2 r_0^2-3 r_0^4\right)^3}T^2\right)\, . \nonumber 
\end{eqnarray}
Note that while doing this expansion we choose to keep the angular momentum $J$ fixed at its extremal value, i.e., $J$ receives no $T$ corrections and it is given by
\begin{equation}
    J = \frac{L r_0^2 \sqrt{\left(L^2-r_0^2\right) \left(L^4+3 L^2 r_0^2\right)}}{\left(L^2-3 r_0^2\right)^2}\, .
\end{equation}
Therefore, we have adopted the canonical ensemble here. However, we should add that this is more of an aesthetic choice than a necessary condition; in fact in appendix \ref{appendix:againKerrAdS4}, we show that the result remains unchanged if we choose to keep $M$ fixed, and let $J$ depend on the temperature. Note that the leading correction to the mass occurs at $O\left(T^2\right)$. This observation is at the heart of the original discussion of the breakdown of thermodynamics at low temperatures in \cite{Preskill:1991tb}. Recent works \cite{Banerjee:2023quv, Kapec:2023ruw, Rakic:2023vhv} clarify important quantum corrections to this dependence and resolve the breakdown of thermodynamics for rotating black holes.

Taking inspiration from  \cite{Banerjee:2023quv, Kapec:2023ruw, Rakic:2023vhv} we introduce the following change of coordinates:
we do a `Bardeen-Horowitz' like scaling transformation $\lbrace \left(r,t,\theta, \phi \right) \rightarrow \left(y, \tau, \theta, \varphi \right) \rbrace$ to obtain for the near-horizon geometry of the near-extremal Kerr-AdS$_4$ black hole \cite{Bardeen:1999px}. The transformation is given by
\begin{align}\label{eq:nhcoordKAdS4_1}
    r &= r_+\left(T\right) + \frac{4 \pi  L^2 r_0^2 \left(L^2+r_0^2\right) }{L^4+6 L^2 r_0^2-3 r_0^4}\, T \left(y-1\right)\, ,\\
    t &= -\frac{i\tau}{2 \pi T } \, , \qquad \theta = \theta \, , \label{eq:nhcoordKAdS4_2}\\
    \phi &= \varphi - i\tau \left(\frac{\left(L^2-3 r_0^2\right) \sqrt{\frac{L^4+3 L^2 r_0^2}{L^2-r_0^2}}}{4 \pi  L^3 r_0}\frac{1}{T} - 1\right).\label{eq:nhcoordKAdS4_3}
\end{align}
 This allows us to simultaneously zoom into the near-horizon region of the black hole and also take a small temperature limit. 
\iffalse

Let us consider the following limit, as described in equation (2.8) of \cite{Kapec:2023ruw}:

\bea
\hat{t}=\frac{2r_0}{\varepsilon(T)}\, t, \qquad 
\hat{r}= r_{+}(T)+r_0\varepsilon(T)(\cosh \eta -1), \qquad \hat{\phi}=\phi +\frac{t}{\varepsilon(T)}-t, \qquad \varepsilon(T)=4\pi r_0 T. \nonumber \\ 
\label{coordinatetrans}
\eea
The expression  $r_+(T)$ contains temperature dependent terms.. However, we can deal with this one problem. 
We want to generalize the above  equation/limit in the case where $r_{+}$ is the outer horizon of our solutions and  $r_0$ is the extremal radius. Note that implementing this limit says nothing about the {\it fate} of the charges or angular momentum. We should compute this, and later, {\color{blue} clarify the ensemble} in which it makes sense. 

\fi
At leading order, the above transformation gives us
\begin{align}  
O(1):  \, \, ds_{(0)}^2 &= \bar g_{\mu\nu}dx^\mu dx^\nu \nonumber \\
&= g_1 (\theta) \left( \frac{dy^2}{y^2-1} + \left(y^2-1\right) d\tau^2 \right) + g_2 (\theta) d\theta^2 + g_3 (\theta)\left(d\varphi + i g_4(y) \left(y-1\right) d \tau \right)^2, \label{eq:zerothKerrAdSmetric}
\end{align}
where 
\begin{align}
    g_1 ( \theta ) &= \frac{L^2 r_0^2 \left(L^2 - r_0^2 + \left(L^2+3 r_0^2\right) \cos^2\theta \right)}{L^4+6 L^2 r_0^2-3 r_0^4}, \\
    g_2 ( \theta ) &= \frac{2 \left(L^6-L^2 r_0^4\right)}{2 L^4 - 3 L^2 r_0^2-3 r_0^4 - r_0^2 \left(L^2+3 r_0^2\right) \cos{2 \theta}}-L^2, \\
    g_3 ( \theta ) &= \frac{4 L^2 r_0^2 \sin^2\theta \left(L^4 - L^2 r_0^2 - r_0^2 \left(L^2+3 r_0^2\right)  \cos^2\theta \right)}{\left(L^2-3 r_0^2\right)^2 \left(L^2 - r_0^2 + \left(L^2+3 r_0^2\right) \cos^2\theta\right)}, \\
    g_4 (y) &= \frac{L^5+6 L^3 r_0^2 - 3 L r_0^4 - \sqrt{\frac{L^4+3 L^2 r_0^2}{L^2-r_0^2}} \left(L^4+4 L^2 r_0^2-3 r_0^4\right) y}{L (y-1) \left(L^4+6 L^2 r_0^2-3 r_0^4\right)}.
\end{align}
%
\iffalse
\begin{eqnarray} 
O(T^0):  \, \, ds_{(0)}^2 &=&\bar g_{\mu\nu}dx^\mu dx^\nu= g_1 (\theta) \left(d\eta^2 + \sinh^2 \eta d\tau^2\right) + g_2 (\theta) d\theta^2 + g_3 (\theta)\left(d\phi -g_4(\eta) d \tau \right)^2\,,   \nonumber \\
\label{eq:zerothKerrAdSmetric}
\end{eqnarray} where 
\begin{align}
    g_1 ( \theta ) &= \frac{L^2 r_0^2 \left(\cos^2\theta \left(L^2+3 r_0^2\right)+L^2-r_0^2\right)}{L^4+6 L^2 r_0^2-3 r_0^4}, \\
    %
    g_2 ( \theta ) &= \frac{2 \left(L^6-L^2 r_0^4\right)}{2 L^4-r_0^2 \cos{2 \theta} \left(L^2+3 r_0^2\right)-3 L^2 r_0^2-3 r_0^4}-L^2, \\
    %
    g_3 ( \theta ) &= \frac{4 L^2 r_0^2 \sin^2\theta \left(L^4-r_0^2 \cos^2\theta \left(L^2+3 r_0^2\right)-L^2 r_0^2\right)}{\left(L^2-3 r_0^2\right)^2 \left(\cos^2\theta \left(L^2+3 r_0^2\right)+L^2-r_0^2\right)}, \\
    %
    g_4 (\eta) &= \frac{\cosh{\eta} \sqrt{\frac{L^4+3 L^2 r_0^2}{L^2-r_0^2}} \left(L^4-4 L^2 r_0^2+3 r_0^4\right)}{L^5+6 L^3 r_0^2-3 L r_0^4}- 1.
\end{align}
\fi
%
This geometry is the analogue of the near-horizon extreme Kerr (\textit{NHEK}) spacetime \cite{Bardeen:1999px, Kapec:2023ruw, Rakic:2023vhv} and reduces to the same in the large AdS$_4$ radius $L \to \infty$ limit.  It may further be checked the geometry in equation \eqref{eq:zerothKerrAdSmetric} has a Ricci scalar $R= - \frac{12}{L^2}$. We shall call this geometry \textit{NHEK-AdS}.

By plugging in the expressions for $M$ and $a$ from equations \eqref{eq:MseriesKerrAdS} and \eqref{eq:aseriesKerrAdS}, we also obtain the $O(T)$ correction to the metric in \eqref{eq:zerothKerrAdSmetric} given by
\begin{eqnarray}
O(T): \,\, \, \,ds_{(1)}^2  &=& T g^{\left(1\right)}_{\mu \nu} dx^\mu dx^\nu\, .
\end{eqnarray}
The expressions for $g^{(1)}_{\mu\nu}$ are typically too lengthy to be presented and not particularly illuminating. 

This concludes our section on near horizon near extremal Kerr geometries and its correction. We now turn to the problem of computing the $\log T$ corrections, arising from the zero modes of the Lichnerowicz operator. 
%
%%%%%%%%%%%%%%%%%%%%%%%%%%%%%%%%%%%%%%%%%%%%%%%%%%%%%
\subsection{Quantum corrections to the entropy and zero modes} 

We are interested in the quantum corrections to the entropy which scale as the logarithm in temperature. This requires a path integral over the massless modes propagating in the near horizon throat. We perform this integral by doing a saddle point analysis, where the saddle is given by the \textit{NHEK-AdS} geometry in \eqref{eq:zerothKerrAdSmetric}. Schematically, we consider $g_{\mu\nu}=\bar{g}_{\mu\nu} + h_{\mu\nu}$, for a normalizable perturbation $h$, and evaluate the 1-loop partition function around a given saddle point $\bar{g}$
\begin{eqnarray} \label{eq:partition function}
    Z = \exp(-I(\bar g)) \int D[h] \exp \Bigg[ -\frac{1}{64\pi} \int d^4x \sqrt{\bar g } ~ \tilde h^{\mu\nu}\Delta^{\Lambda} h_{\mu\nu}\Bigg]\, ,
\end{eqnarray} with $\tilde h ^{\mu\nu} = h^{\mu\nu} - \frac{1}{2}\bar{g}^{\mu\nu} h_{\mu\nu}$. Note that here we use the gauge fixed action which satisfies the harmonic gauge condition given by
\begin{eqnarray}
    \mathcal{L}_{GF} = \frac{1}{32 \pi} \bar g_{\mu\nu} \left(\bar \nabla_\alpha h ^{\alpha \mu} - \frac{1}{2} \bar \nabla^\mu h^\alpha _{~ \alpha }  \right) \left(\bar \nabla_\beta h ^{\beta \nu} - \frac{1}{2} \bar \nabla^\nu h^\beta _{~ \beta }  \right)\, .
\end{eqnarray}
For this gauge fixed action, the linearized kinetic operator for $h_{\mu\nu}$, also known as the Lichnerowicz operator, is given by~\cite{Christensen:1979iy}
\begin{eqnarray}
    \Delta^\Lambda h_{\mu\nu} = \frac{1}{32 \pi} \Bigg(-\bar \nabla^2  h_{\mu\nu} + 2 \bar R _{\mu\rho} h^{\rho}_{~\nu} - 2 \bar  R_{\mu\rho\nu\sigma} h^{\rho\sigma} - 4 \left(\bar R_{\nu\sigma }- \frac{1}{4} \bar g_{\nu\sigma} \bar R\right) h^\sigma _{~\mu } - 2 \Lambda h_{\mu\nu}\Bigg)\, . \qquad 
\end{eqnarray}
As we shall see in section \ref{sec:logTAdS}, the $\log T$ corrections arise from the zero modes of the Lichnerowicz operator. In the remaining part of this section we will evaluate the zero modes of the same, anticipating the arguments given in next section. Armed with the expressions for the zero modes, we would then be ready to directly evaluate the $\log T$ correction in section~\ref{sec:logTAdS}.

Typically, the Lichnerowicz operator is quite intractable, more so in our case where the metric is not Ricci flat. We would need a judicious ansatz for the vector field generating the zero modes as diffeomorpshisms. Following the prescription as given in \cite{Rakic:2023vhv}, we choose an ansatz for the vector field as 
\begin{eqnarray}
    \xi^{(n)} = e^{i n \tau} \left(f_1 (y) \frac{\partial}{\partial y } + f_2 (y) \frac{\partial}{\partial \tau} + f_3 (y) \frac{\partial}{\partial \varphi }  \right)\,  \qquad 
\end{eqnarray}  and we  generate diffeomorphisms given by $h_{\mu\nu} = \mathcal{L}_{\xi} g^{(0)}_{\mu\nu}$. The goal now is to find the functions $f_{1,2,3} (y)$ such that the $h_{\mu\nu}$ generated belongs to the kernel of the Lichnerowicz opertor. We state the result first. The vector fields are given by (for $|n| \geq 2$) 
\begin{align} \label{eq:vectormodesKerrAdS}
\begin{split}
    &\xi^{(n)} =  \frac{i e^{i n \tau} \left(\frac{y-1}{y+1}\right)^{\frac{|n|-2}{2}} \left(n^2+|n| y+y^2-1\right)}{2 |n| \left(n^2-1\right) (y+1)^2} \partial_{\tau} + \frac{e^{i n \tau}\left(\frac{y-1}{y+1}\right)^{\frac{|n|}{2}} (|n|+y)}{2 \left(n^2-1\right)} \partial_{y} + \\ & \frac{\begin{pmatrix}
        e^{i n \tau} \left(\frac{y-1}{y+1}\right)^{\frac{|n|}{2}} \Bigg(L^5 \left(n^2 y + |n|\right)-L^4 \sqrt{\frac{L^4+3 L^2 r_0^2}{L^2-r_0^2}} \left(n^2+|n| y+y^2-1\right) - \\
        6 L^2 r_0^2 \sqrt{\frac{L^4+3 L^2 r_0^2}{L^2-r_0^2}} \left(n^2+|n| y+y^2-1\right)+3 r_0^4 \sqrt{\frac{L^4+3 L^2 r_0^2}{L^2-r_0^2}} \left(n^2+|n| y+y^2-1\right)-9 L r_0^4 (n^2 y + |n|)\Bigg)
    \end{pmatrix}}{2 |n| \left(n^2-1\right) \left(y^2-1\right) \sqrt{\frac{L^4+3 L^2 r_0^2}{L^2-r_0^2}} \left(L^4+6 L^2 r_0^2-3 r_0^4\right)} \partial_{\varphi} 
\end{split}
\end{align}
 such that the diffeomorphism generated by this vector field (given in equation \eqref{eq:zeromodesKerrAdS}) belongs to the kernel of the Lichnerowicz operator. Upon introducing a new radial coordinate $y=\cosh{\eta}$, we may also write the above as
 \begin{equation}
 \begin{split} \label{eq:vectormodesKerrAdS:new_patch}
     &\xi^{(n)} = \frac{e^{i n \tau} \csch^2{\eta} \tanh^{|n|}\left(\frac{\eta}{2}\right)}{2 |n| \left(n^2-1\right)} \Bigg( i \left( n^2+\cosh\eta \left(\cosh\eta+|n|\right)-1 \right) \partial_{\tau} + \left(|n| \sinh{\eta} \left( \cosh\eta + |n| \right)\right) \partial_{\eta} + \\ & \frac{L \left(L^4 - 9 r_0^4 \right)  \left( n^2 \cosh{\eta} + |n| \right) - \left( L^4 + 6L^2r_0^2 - 3 r_0^4 \right) \sqrt{\frac{L^4+3 L^2 r_0^2}{L^2-r_0^2}} \left( n^2+\cosh {\eta} \left( \cosh{\eta} + |n| \right) - 1 \right)}{\sqrt{\frac{L^4+3 L^2 r_0^2}{L^2-r_0^2}} \left(L^4+6 L^2 r_0^2-3 r_0^4\right)} \partial_{\varphi} \Bigg)\, .
\end{split}
 \end{equation}

Typically, it becomes an involved problem to solve directly for the functions $f_i\left(y\right)$. Borrowing inspiration from the treatment of zero modes in \cite{Rakic:2023vhv}, we illustrate below some of the steps to solve for the vector field. 
%%%%%%%%%%%%%%%%%%%%%%%%%%%
We act the Lichnerowicz operator ($\Delta^\Lambda$) on the spin 2 field $h_{\mu \nu} = 2\nabla_{(\mu} \xi_{\nu)}$ which represents the change of the \textit{NHEK-AdS} metric under the diffeomorphism generated by the vector field $\xi$. Specifically, we observe that
\begin{eqnarray}
    (\Delta^{\Lambda} h)_{\theta\theta}\Big|_{\theta=0}^{\theta=\frac{\pi}{2}} \sim \left(i n f_2 (y)+f^\prime_1 (y)\right)\, 
\end{eqnarray} where the proportionality factor is a complicated function of the metric functions, but is independent of any other $f_i$. This immediately sets
\begin{eqnarray}
    f_2 (y) = \frac{i f_1^\prime (y)}{n}\, . 
\end{eqnarray} Next, we choose an ansatz similar to \cite{Rakic:2023vhv} for $f_3 (y)$ and proceed. That is, we choose
\begin{eqnarray}
    f_3(y)=\frac{(y-1) f^\prime_1(y) -f_1(y)}{n}+\delta f_3(y)
\end{eqnarray} and solve for $\delta f_3(y)$ from the equation
\begin{eqnarray}
   \partial_\theta ( \Delta^{\Lambda} h_{\theta y})\Big|_{\theta=0} +\frac{1}{2} \partial_\theta ( \Delta^{\Lambda} h_{\theta y})\Big|_{\theta=\frac{\pi}{2}} = 0. 
\end{eqnarray} 
Here, we see a difference with the Kerr case \cite{Rakic:2023vhv}, and we do \textit{not} obtain $\delta f_3 = 0$. Rather, we get a complicated function that we do not write here for brevity. We also do a consistency check which shows that $\delta f_3$ vanishes as $L\rightarrow \infty$, as expected. We are ultimately left with a single undetermined function $f_1\left(y\right)$, which can be solved by brute force. Finally, we arrive at 
\begin{eqnarray}
    f_1 (y) = \left(\frac{y-1}{y+1}\right)^{\frac{|n|}{2}} \frac{ (|n|+y)}{2 \left(n^2-1\right)}\, , \quad |n| \geq 2.
\end{eqnarray} This is exactly the solution found by the authors in \cite{Rakic:2023vhv}. Once we have found the functions $f_i$'s, the family of normalized zero modes of the  Lichnerowicz operator is obtained as 
\begin{equation} \label{eq:zeromodesKerrAdS}
\begin{split}
    h^{(n)}_{\mu\nu}dx^{\mu}dx^{\nu} &= \sqrt{\frac{3 |n| \left(n^2-1\right)\left(L^6 + 3L^4 r_0^2 - 21L^2 r_0^4 + 9r_0^6\right)}{64 \pi^2 L}} \frac{L^2 - r_0^2 + \left(L^2 + 3 r_0^2 \right) \cos^2{\theta}}{L^4 + 6 L^2 r_0^2 - 3 r_0^4} \\ & \qquad e^{i n \tau} \left( \frac{y-1}{y+1} \right)^{\frac{|n|}{2}} \left(-d\tau^2 + 2i \frac{|n|}{n} \frac{d\tau dy}{y^2 - 1} + \frac{dy^2}{\left(y^2-1\right)^2} \right) \,, \quad |n| \geq 2\,\, . 
    \end{split}
\end{equation}  One can immediately check that 
\begin{eqnarray} \label{eq:kernelofLich}
     (\Delta^{\Lambda} h)_{\mu\nu} = 0\, , \qquad {\mu,\nu} \in \{y, \tau,  \theta, \varphi\}.  
\end{eqnarray}

%%%%%%%%%%%%%%%%%%%%%
The zero modes in equation \eqref{eq:zeromodesKerrAdS} are generated by diffeomorphisms $\tensor{h}{^{(n)}_{\mu\nu}} \propto \mathcal{L}_{\xi} \tensor{\bar{g}}{_{\mu\nu}}$, where the vector fields are given by \eqref{eq:vectormodesKerrAdS}. 
 Note that they reduce to the same set of vector modes as is reported in~\cite{Rakic:2023vhv} in the large $L$ limit.

\paragraph{Large Diffeomorphism:} At the very onset, we should point that the diffeomorphisms generated by the vector field $\xi$ are large gauge transformations and they can not be gauged away. To see this, we first note that the  the vector field $\xi$ does not die off at the asymptotic boundary of AdS$_2$. Specifically from \eqref{eq:vectormodesKerrAdS:new_patch}, we see that 
\begin{equation} \label{eq:diffatboundary}
    \xi^{(n)}\big|_{\eta\to\infty} = \frac{e^{i n \tau}}{2 |n| \left(n^2 - 1\right)}\left( |n| \partial_{\eta} + i\partial_{\tau}-\partial_\varphi \right)\,,
\end{equation} i.e., the vector fields themselves are $O(1)$ near the boundary. However, the zero modes $\{h_{\mu\nu}^{(n)}\}$ themselves are normalizable and are part of the physical spectrum and need to be integrated over in the path integral. We use the normalization \cite{Camporesi:1994ga}
\begin{equation}
  \langle h, h \rangle =   \int d^4x \sqrt{\bar{g}} \left(h^{(n)}\right)^{\ast}_{\mu\nu} \left(h^{(n)}\right)^{\mu\nu}
\end{equation} as would be appropriate for the space of $L_2$ integrable functions necessary to define path integrals. Setting the above inner product one, i.e., 
\begin{eqnarray}
     \int d^4x \sqrt{\bar{g}} \left(h^{(n)}\right)^{\ast}_{\mu\nu} \left(h^{(n)}\right)^{\mu\nu} = 1
\end{eqnarray} fixes the coefficient for $h^{(n)}_{\mu\nu}dx^{\mu}dx^{\nu}$ in equation \eqref{eq:zeromodesKerrAdS}.

A bit more insight might be had from the behaviour of the diffeomorphism generating vector fields \eqref{eq:vectormodesKerrAdS} at the boundary. In the coordinates $\{\eta, \tau, \theta, \varphi \}$
\begin{eqnarray} \label{eq:Schwarzian action at boundary}
    \xi \sim \varepsilon^\prime(\tau) \partial_\eta - \varepsilon(\tau) \partial_\tau -i \varepsilon(\tau) \partial_{\varphi}\, , \quad  \varepsilon(\tau) = \sum_{n=2}^{\infty} f_n  \exp (i n \tau)\, . 
\end{eqnarray}
The vector field then generates the reparametrization $\eta \rightarrow \eta + \varepsilon^\prime (\tau)\,, \tau \rightarrow \tau - \varepsilon,  \varphi\rightarrow \varphi -i\varepsilon$. Following the arguments presented in \cite{Iliesiu:2022onk, Maldacena:2016upp}, we conclude that the vector fields generating the large diffeomorphisms are the Schwarzian modes which act on the boundary via \eqref{eq:Schwarzian action at boundary}.
%%%%%%%%%%%%%%%%%%%%%%%%%%%%%%%%%%%%%%%%%%%%%%%%%%
\subsection{Log $T$ corrections from zero modes} \label{sec:logTAdS}

 In this section, we motivate the claim made in the last section that only the zero modes of the Lichnerowicz operator contribute to $\log T$ corrections. The claim is based on perturbation theory. Then, building on that claim, we actually evaluate the $\log T$ correction, where the final result is presented in equation \eqref{eq:final result _ Kerr AdS}. 
 
 First, we note that the correction to metric \eqref{eq:zerothKerrAdSmetric} induces a correction $\delta \Delta^{\Lambda}$ to the Lichnerowicz operator . In order to find the eigenvalue of the corrected Lichnerowicz operator ($\Delta^\Lambda+\delta \Delta^\Lambda$), we take help of perturbation theory. At first order, we have 
  \begin{equation}
        \left(\Delta^{\Lambda} + \delta \Delta^{\Lambda}\right)\left(h_n + \delta h_n\right) = \left(\lambda_n^{0} + \delta \lambda_n\right)\left(h_n + \delta h_n\right),
    \end{equation}  where $\{h_n,\lambda^0_n\}$ is the eigenspectrum and $\{\delta h_n,\delta \lambda_n\}$ is the respective correction induced. Then, at first order, we have  
     \begin{equation} 
        \delta\lambda_n = \int d^4x \sqrt{\bar{g}}\: \left(h^{(n)}\right)_{\alpha\beta} \left(\delta \Delta^{\Lambda}\right)^{\alpha\beta, \mu\nu} \left(h^{(n)}\right)_{\mu\nu},
        \label{eqn: LiftedEigenvalue}
    \end{equation}

    Since, the partition function in \eqref{eq:partition function} is a Gaussian integral over bosonic fields, ignoring overall factor, which play little role, we have
    \begin{eqnarray} \label{eq:generalexpZKerrAdS}
        Z \sim \prod_{n} \frac{1}{\left(\lambda^0_n + \delta \lambda_n  (T)\right)}\, , \qquad \log Z \sim - \sum_n \log \left(\lambda^0_n + \delta \lambda_n  (T) \right)\, , 
    \end{eqnarray} To proceed further, we need two crucial observations. We observe that  $\lambda_n^0$ is $T$-independent, being the eigenvalue of the Lichnerowicz operator evaluated at $O(1)$ and therefore, $\log T$ corrections would necessarily come from the corrected eigenvalue $\delta\lambda_n  (T)$. Furthermore, for non-zero $a$, we note that the Maclaurin expansion of $\log (a + x(T))$ is always a polynomial series in small~$x\left(T\right)$. Specifically, we mean that
    \begin{eqnarray}
        \log (a + x(T)) = \log a + \sum^\infty_{n=1} \frac{x^n(T)}{a^n n}\, . 
    \end{eqnarray} Therefore, corrections of the form $\log \left( x(T) \right)$ come only when $a=0$. Adapting this observation to our case, this implies that the $\log T $ corrections are obtained only from the zero modes of the Lichnerowicz operator when $\lambda^0_n=0$. This establishes our claim in the last section that for the $\log T$ correction, it suffices to focus on the zero modes of the Lichnerowicz operator. Therefore, we are left with the job of evaluating the corrected eigenvalue of the zero modes, via equation \eqref{eqn: LiftedEigenvalue}. The integral is quite involved. However, it can be evaluated and we quote the result here directly. We have
        \begin{equation} \label{eq:correctedeigenval_KerrAdS}
    \delta\lambda_n = \frac{3 n \left(1 - \frac{r_{0}^2}{L^2} \right) T}{64\,r_{0}}, \quad n \geq 2 \, .
\end{equation}
    Therefore, setting $\lambda^0_n=0$ in equation \eqref{eq:generalexpZKerrAdS} we obtain the $\log T$ corrections as
\begin{eqnarray} \label{eq:logTKerrAdS}
    \delta \log Z \Bigg|_{\log T} = -\log \Bigg[\prod^\infty_{n=2} \left(\frac{3 n \left(1 - \frac{r_{0}^2}{L^2} \right) T}{64\,r_{0}}\right) \Bigg]= \log \Bigg[\prod^\infty_{n=2} \left(\frac{64\,r_{0}}{3 n \left(1 - \frac{r_{0}^2}{L^2} \right) T}\right) \Bigg] . \qquad 
\end{eqnarray} The infinite product above can be regularized using standard zeta function regularization \cite{GonzalezLezcano:2023cuh} given by
\begin{align}
    \prod_{n=2}^\infty \frac{x}{n} \sim x^{-\frac{3}{2}}.
\end{align} Setting $x = T^{-1}$ and ignoring constants irrelevant to the present pursuit, we obtain
\begin{eqnarray} \label{eq:final result _ Kerr AdS}
    \delta \log Z\Bigg|_{\log T} = \frac{3}{2} \log \left(\frac{T}{T_q}\right) + O(1), \quad T_q \equiv \frac{64\,r_0}{3\left(1-\frac{r_0^2}{L^2}\right)}\,.
\end{eqnarray}
The constant indicated by $O(1)$ depends on the scheme. In this manuscript we will use $T_q$ as the temperature scale below which the logarithmic correction become dominant.

\noindent Two final comments are in order here.
\begin{enumerate}
    \item  Note that on taking the large $L$ limit in \eqref{eq:logTKerrAdS}, one recovers the result for the Kerr black hole in flat space presented in \cite{Kapec:2023ruw,Rakic:2023vhv}.
    \item The result agrees with the generic prediction of \cite{Rakic:2023vhv}. We have derived it explicitly, and we confirm the independence of the $\log T$ correction with respect to the ensemble (canonical versus grand canonical). For the grand canonical ensemble (where $m$ is kept fixed), see appendix \ref{appendix:againKerrAdS4}. 
\end{enumerate}

%%%%%%%%%%%%%%%%%%%%%%%%%%%%%%%%%%%%%%%%%%%%%%%%%%%
\section{Kerr-Newman-AdS$_4$}\label{Sec:KN-AdS}
In this section, we provide the details of treating the path integral for the near-extremal Kerr-Newman-AdS black hole in four dimensions. We highlight a cancellation pattern among terms in the Lichnerowicz operator. Such cancellations are verified in all the explicit cases we considered in this paper. We start with the Einstein-Maxwell action,
\be
 I = -\frac{1}{16\pi} \int_{\mathcal{M}} d^4x \sqrt{-g} \left(R - 2\Lambda -F^2\right)+I_{boundary}.
\ee
The Kerr-Newman-AdS$_4$ solution is given by \cite{Caldarelli:1999xj}
\begin{equation} \label{eq:metricKerrNewmanAdS4}
    ds^2 = - \frac{\Delta_r}{\Sigma^2} \left(dt - \frac{a}{\Xi} \sin^2 \theta d\phi\right)^2 + \frac{\Sigma^2}{\Delta_r}dr^2 + \frac{\Sigma^2}{\Delta_\theta}d\theta^2 + \frac{\Delta_\theta}{\Sigma^2}\sin^2 \theta\left( a dt- \frac{\left(r^2+a^2\right)}{\Xi}d\phi \right)^2 \, ,
\end{equation}
where
\begin{eqnarray}
    \Delta_r &=& \left(r^2+a^2\right)\left(a+g^2r^2\right) - 2 M r+q^2 \, , \quad  \Delta_\theta = 1 - g^2a^2 \cos^2\theta\, , \\
    \Xi &=& 1 - g^2a^2 \, ,  \qquad \Sigma = \sqrt{r^2 + a^2 \cos^2 \theta}\, . 
\end{eqnarray}
The cosmological constant $\Lambda$ is related to  $g\equiv\frac{1}{L}$ by $\Lambda = -3g^2$.
The gauge field is given by (we consider only the electric case, and the magnetically charged case can be treated analogously):
\be
A=-\frac{qr}{\Sigma^2}dt+\frac{qra\sin^2\theta}{\Sigma^2 \Xi}d\phi.
\ee
The temperature of the black hole is given by 
\be
\label{eq:KN-AdS_Temperature}
T=\frac{r_{+} \left(1 + g^2 a^2 + 3 g^2 r_{+}^2 - \frac{a^2 + q^2}{r_{+}^2}\right)}{4 \pi (r_{+}^2 + a^2)},
\ee
where the $r_+$ denotes radius of the outer horizon. 

%%%%%%%%%%%%%%%%%%%%%%%%%%%%%%%%%%%%%%%%%%%%%%%%%%%%%%%%
\subsection{Ensemble choice and near-horizon limit}
Under the extremal condition, we have
\bea
m_0&=&r_0 + a^2 g^2 r_0 + 2 g^2 r_0^3,\\
q_0&=&\sqrt{-a^2 + r_0^2 + a^2 g^2 r_0^2 + 3 g^2 r_0^4},
\eea
where $r_0$ is the radius of the extreme black hole horizon. In our ensemble, we choose $a$ and $r_0$ as the two free parameters, $m$ and $r_+$ depend on $T$:
\bea
m(T)=m_0+c_1 T+c_2 T^2+c_3 T^3+O(T^4),\\
r_+(T)=r_0+d_1 T+d_2 T^2+d_3 T^3+O(T^4),
\eea
where the coefficients $c_i,d_i$ can be obtained by solving \eqref{eq:KN-AdS_Temperature} and $\Delta_{r}=0$. 
We verify that $c_1=0$. Other coefficients are lengthy and not illuminating to be written here. The Hawking entropy can be calculated using the above temperature dependence
\be
\label{eq:KN-AdS_Classical Entropy}
S=\frac{\pi (a^2 + r_0^2)}{1 - a^2 g^2} + \frac{4 \pi^2 r_0 (a^2 + r_0^2) }{(1 - a^2 g^2)(1 + a^2 g^2 + 6 g^2 r_0^2)}T+O(T^2).
\ee
To zoom into the near-horizon region, we conduct the following coordinate transformation:
\be
r=r_+(T)+T k_1(y-1),\ t=\frac{-i\tau}{k_2 T},\ \phi=\varphi+\frac{-ik_3\tau}{T}+k_4i\tau,\ 
\ee
where 
\be
\begin{aligned}
k_1 &= \frac{2 \pi (a^2 + r_0^2)}{1 + a^2 g^2 + 6 g^2 r_0^2}, \\
k_2 &= 2 \pi, \\
k_3 &= \frac{-a (-1 + a^2 g^2)}{2 \pi (a^2 + r_0^2)}, \\
k_4 &= \frac{-2 a (-1 + a^2 g^2) r_0}{(a^2 + r_0^2) (1 + a^2 g^2 + 6 g^2 r_0^2)}.
\end{aligned}
\ee
By taking the $T\to0$ limit, we obtain the near-horizon geometry at the zeroth order of $T$,
\be
\label{eq:KN-AdS4_NearHorizon Zeroth}
\Bar{g}_{\mu\nu}dx^\mu dx^\nu =  g_1(\theta)\left(\frac{dy^2}{y^2-1} + d\tau ^2 \left(y^2-1\right)\right)+   g_2(\theta)d\theta^2 +g_{3}( \theta) \Big(d\varphi+i k(y-1)  d\tau   \Big)^2,
\ee
where
\be
\begin{aligned}
g_1(\theta)=&\frac{r_0^2 + a^2 \cos^2\theta}{1 + a^2 g^2 + 6 g^2 r_0^2},\\
g_2(\theta)=&\frac{r_0^2 + a^2 \cos^2\theta}{1 - a^2 g^2 \cos^2\theta},\\
g_3(\theta)=&-\frac{(a^2 + r_0^2)^2 (-1 + a^2 g^2 \cos^2 \theta) \sin^2 \theta}{(-1 + a^2 g^2)^2 (r_0^2 + a^2 \cos^2 \theta)},\\
k=&-k_4.\\
\end{aligned}
\ee
By keeping $O(T)$ when $T\to0$, we also obtain $\delta g_{\mu\nu}$, the near-horizon geometry at $O(T)$. This expression is lengthy and not very illuminating to be written here.
%%%%%%%%%%%%%%%%%%%%%%%%%%%%%%%%%%%%%%%%%%%%%%%%%%%
\subsection{Lichnerowicz operator and the zero modes}
The Lichnerowicz operator of our Einstein-Maxwell action with a  negative cosmological constant is
\bea
h_{\alpha\beta}\Delta^{\alpha \beta, \mu \nu}_{L}h_{\mu\nu}=h_{\alpha\beta}(\Delta^{\alpha \beta, \mu \nu}_{EH}-2\Lambda\Delta^{\alpha \beta, \mu \nu}_{1}-2\Delta^{\alpha \beta, \mu \nu}_{F})h_{\mu\nu},
\label{eq:KN-AdS4 Linch operator}
\eea
where $\Delta^{\alpha \beta, \mu \nu}_{EH},\Delta^{\alpha \beta, \mu \nu}_{1},\Delta^{\alpha \beta, \mu \nu}_{F}$ can be found in appendix \ref{appendix:Lich operator}.
At zero temperature, i.e. when $g=\bar{g}$, one can verify that the above operator has the following zero modes
\bea
\notag
h_{\mu\nu}^{(n)}dx^\mu dx^\nu=&c_n& e^{i n \tau} \left(\frac{-1 + y}{1 + y}\right)^{\frac{\abs{n}}{2}} \left(a^2 + 2 r_0^2 + a^2 \cos(2 \theta)
\right)(d\tau^2-2i\frac{\abs{n}}{n}\frac{d\tau dy}{y^2-1}-\frac{dy^2}{(y^2-1)^2}),\\
&&\abs{n}\geq2,
\label{eq:KN-AdS4 zero modes}
\eea

where $c_n$ is the normalization constant.
%%%%%%%%%%%%%%%%%%%%%%%%%%%%%%%%%%%%%%%%%%%%%%%%%%%
\subsection{The lifted eigenvalues and entropy correction}
By turning on a small temperature, the ``zero modes'' will be lifted and gain a nonzero eigenvalue. Now we substitute $g=\bar{g}+\delta g$ into the Lichnerowicz operator \eqref{eq:KN-AdS4 Linch operator} and apply it to the zero modes \eqref{eq:KN-AdS4 zero modes}. The lifted eigenvalue is
\be
\delta\lambda_n=\int dx^4 \sqrt{\bar{g}}h^{(n)*}_{\alpha\beta}\delta\Delta_L^{\alpha\beta,\mu\nu} h^{(n)}_{\mu\nu}.
\ee
The expression of the operator is intractable, but the integration is straight forward and the result is simple.
\be
\delta\lambda_n=\frac{3 n r_0T}{8 (a^2 + 3 r_0^2)},\quad   \abs{n}\geq2.
\ee
The contribution of the extremal zero modes to the low-temperature partition function is therefore
\be
\delta \log Z=\log(\prod_{n\geq2}\frac{\pi}{\delta\lambda_n})=\frac{3}{2}\log (\frac{T}{T_q})+O(1).
\ee
The $O(1)$ term is not reliable. $T_q$ is a temperature scale under which the quantum correction should be considered. We use the inverse of the $O(T)$ coefficient of the classical entropy \eqref{eq:KN-AdS_Classical Entropy} as this scale
\be
T_q=\frac{(1 - a^2 g^2) (1 + a^2 g^2 + 6 g^2 r_0^2)}{4\pi^2r_0 (a^2 + r_0^2)}.
\ee
This is the temperature we expect the classical low temperature thermodynamics to break down. 
%%%%%%%%%%%%%%%%%%%%%%%%%%%%%%%%%%%%%%%%%%%%%%%%%%%
\subsection{Cancellations within the Lichnerowicz operator}
If calculating every term in the Lichnerowicz operator respectively, one can find a pattern that hold in all cases considered in this paper: only two terms really contribute to the lifted eigenvalue. The other terms always cancel with each other. In this subsection, we use the example of Kerr-Newman-AdS$_4$ black hole to illustrate this cancellation. For convenience, we first rename the terms that appear in the Lichnerowicz operator \eqref{eq:KN-AdS4 Linch operator},
\begin{align}
{\rm Term\, 1a:}& \quad h_{\alpha\beta}\ \delta(\frac{1}{2}g^{\alpha\mu}g^{\beta\nu}\square)h_{\mu\nu}  \nonumber \\
{\rm Term\, 1b:}& \quad   h_{\alpha\beta}\ \delta(-\frac{1}{4}g^{\alpha\beta}g^{\mu\nu}\square )h_{\mu\nu}=0\nonumber \\
{\rm Term\, 2:}& \quad   h_{\alpha\beta}\ \delta(R^{\alpha\mu\beta\nu})h_{\mu\nu}\nonumber \\
{\rm Term\, 3a:}& \quad  h_{\alpha\beta}\ \delta(R^{\alpha\mu}g^{\beta\nu})h_{\mu\nu}
\nonumber \\
{\rm Term\, 3b:}& \quad  h_{\alpha\beta}\ \delta(-R^{\alpha\beta}g^{\mu\nu})h_{\mu\nu}=0 \nonumber \\
{\rm Term\, 4a:}& \quad  h_{\alpha\beta}\ \delta(-\frac{1}{2}R g^{\alpha \mu}g^{\beta\nu} )h_{\mu\nu}
\nonumber \\
{\rm Term\, 4b:}& \quad  h_{\alpha\beta}\ \delta(\frac{1}{4}R g^{\alpha\beta}g^{\mu\nu})h_{\mu\nu} =0 
 \nonumber \\
 {\rm Term\, 5a:}& \quad  h_{\alpha\beta}\ \delta(\Lambda g^{\alpha \mu}g^{\beta\nu})h_{\mu\nu} \nonumber \\
{\rm Term\, 5b:}& \quad  h_{\alpha\beta}\ \delta(-\frac{1}{2}\Lambda g^{\alpha \beta}g^{\mu\nu})h_{\mu\nu} =0\nonumber \\
{\rm Term\, 6a:}& \quad  h_{\alpha\beta}\ \delta(\frac{1}{2} F^2 g^{\alpha \mu} g^{\beta \nu})h_{\mu\nu}
\nonumber \\
{\rm Term\, 6b:}& \quad  h_{\alpha\beta}\ \delta(-\frac{1}{4} F^2g^{\alpha \beta} g^{\mu \nu})h_{\mu\nu} =0\nonumber \\
{\rm Term\, 7:}& \quad  h_{\alpha\beta}\ \delta(-2F^{\alpha \mu} F^{\beta \nu})h_{\mu\nu} \nonumber \\
{\rm Term\, 8a:}& \quad  h_{\alpha\beta}\ \delta(-4 F^{\alpha \gamma} F_{\ \gamma}^\mu g^{\beta \nu})h_{\mu\nu} \nonumber \\
{\rm Term\, 8b:}& \quad  h_{\alpha\beta}\ \delta(2F^{\alpha \gamma} F_{\ \gamma}^\beta g^{\mu \nu})h_{\mu\nu} =0 \, ,\label{eq:cancelling terms KN-AdS4}
\end{align}
where $\delta$ means the above expressions are of order $T$. As $h_{\mu\nu}$ are zero modes, the $O(1)$ results will cancel with each other automatically. We find two patterns for the above terms,
\begin{itemize}
    \item As indicated in \eqref{eq:cancelling terms KN-AdS4}, all terms containing $g^{\mu\nu}h_{\mu\nu}$ are zero by themselves.
    \item ${\rm Term \, 3a}+{\rm Term\, 4a}+{\rm Term \, 5a}+{\rm Term \, 6a}+{\rm Term \, 7}+{\rm Term \, 8a}=0.$ 
\end{itemize}
Therefore, only two terms contribute non-trivially
\be
\label{eq:Cancellation}
h_{\alpha\beta}\Delta^{\alpha \beta, \mu \nu}_{L}h_{\mu\nu}={\rm Term \, 1a}+{\rm Term\, 2}= h_{\alpha\beta}\ \delta(\frac{1}{2}g^{\alpha\mu}g^{\beta\nu}\square+R^{\alpha\mu\beta\nu})h_{\mu\nu}.
\ee
It is noteworthy that the result of \eqref{eq:Cancellation} can be verified for all cases included in this paper, including the upcoming supersymmetric case and five-dimensional cases in the corresponding next sections. For the supersymmetric case in section \ref{Sec:Sugra}, we have two gauge fields and two additional scalar fields. There are more terms in the Lichnerowicz operator, but only the two terms indicated in \eqref{eq:Cancellation} contribute non-trivially. This cancellation naturally reduces to previous studies of asymptotically flat Kerr and Kerr-Newman black holes in four dimensions \cite{Kapec:2023ruw,Rakic:2023vhv}.
% In fact, we have explored this cancellations beyond the explicit examples considered in this manuscript. In particular, we have confirmed it for Einstein-Maxwell with scalar fields and negative cosmological constant. We will discuss the generality elsewhere.

%%%%%%%%%%%%%%%%%%%%%%%%%%%%%%%%%%%%%%%%%%%%%%%%%%%
\section{Kerr-Newmann-AdS black hole in  ${\cal N}=4$ gauged supergravity}\label{Sec:Sugra}

The 4d $\mathcal{N}=4$ gauged supergravity can be obtained by the truncation of the 11d supergravity \cite{Chong:2004na}, here we exclusively focus on its bosonic sector which is given by
\begin{align}
    \begin{split}
        \mathcal{L}_{4}=& R * 1-\frac{1}{2} * \mathrm{d} \zeta \wedge \mathrm{d} \zeta-\frac{1}{2} e^{2 \zeta} * \mathrm{d} \chi \wedge \mathrm{d} \chi-\frac{1}{2} e^{-\zeta} * F_{(2) 2} \wedge F_{(2) 2}-\frac{1}{2} \chi F_{(2) 2} \wedge F_{(2) 2} \\
        &-\frac{1}{2\left(1+\chi^{2} e^{2 \zeta}\right)}\left(e^{\zeta} * F_{(2)1} \wedge F_{(2) 1}-e^{2 \zeta} \chi F_{(2) 1} \wedge F_{(2) 1}\right) \\
        &+g^{2}\left(4+2 \cosh \zeta+e^{\zeta} \chi^{2}\right) * 1,
    \end{split}
    \label{eq:sugra action}
\end{align}
where $\zeta$ and $\chi$ are the dilaton and axion. The subscript in parenthesis denotes the degree of the form. The solution has two pairwise equal charges and therefore two gauge potential $A_{(1)1}$ and $A_{(1)2}$.

The non-extremal rotating, electrically  charged asymptotically AdS$_4$ black hole solution with gauge group $U(1) \times U(1)$ in 4d $\mathcal{N}=4$ gauged supergravity was constructed in \cite{Chong:2004na}. The solution is characterized by four parameters $(a, m, \delta_1, \delta_2)$. The metric, the scalars and the gauge fields are given by
\be
  ds^2 = - \frac{\Delta_r}{W} \left(dt - \frac{a\, \textrm{sin}^2 \theta}{\Xi} d\phi \right)^2 + W \left(\frac{dr^2}{\Delta_r} + \frac{d\theta^2}{\Delta_\theta} \right) + \frac{\Delta_\theta\, \textrm{sin}^2 \theta}{W} \left[a\, dt - \frac{r_1 r_2 + a^2}{\Xi} d\phi \right]^2\, ,\label{eq:AdS4Metric1}
\ee
\begin{align}
\begin{split}
  e^{\zeta} & = \frac{r_1^2 + a^2\, \textrm{cos}^2 \theta}{W}\, ,\qquad \chi = \frac{a (r_2 - r_1)\, \textrm{cos}\, \theta}{r_1^2 + a^2\, \textrm{cos}^2 \theta}\, ,\\
  A_1 & = \frac{2 \sqrt{2} m\, \textrm{sinh} (\delta_1)\, \textrm{cosh} (\delta_1)\, r_2}{W} \left(dt - \frac{a\, \textrm{sin}^2 \theta}{\Xi} d\phi \right) + \alpha_{41}\, dt\, ,\\
  A_2 & = \frac{2 \sqrt{2} m\, \textrm{sinh} (\delta_2)\, \textrm{cosh} (\delta_2)\, r_1}{W} \left(dt - \frac{a\, \textrm{sin}^2 \theta}{\Xi} d\phi \right) + \alpha_{42}\, dt\, ,
\end{split}
\end{align}
where
\begin{align}
\begin{split}
  r_i & \equiv r + 2 m\, \textrm{sinh}^2 (\delta_i)\, ,\quad (i = 1, 2)\\
  \Delta_r & \equiv r^2 + a^2 - 2 m r + g^2 r_1 r_2 (r_1 r_2 + a^2)\, ,\\
  \Delta_\theta & \equiv 1 - g^2 a^2\, \textrm{cos}^2 \theta\, ,\\
  W & \equiv r_1 r_2 + a^2\, \textrm{cos}^2 \theta\, ,\\
  \Xi & \equiv 1 - a^2 g^2\, ,
\end{split}
\end{align}
and $g \equiv L^{-1}$ is the inverse of the AdS$_4$ radius. Note that we have added pure gauge terms to the two gauge fields where $\alpha_{41}$ and $\alpha_{42}$ are constant.

The non-extremal asymptotically AdS$_4$ black holes with four degenerate electric charges ($Q_1 = Q_2$, $Q_3 = Q_4$) and one angular momentum $J$ have been found in \cite{Cvetic:2005zi}, which are characterized by four parameters $(a, m, \delta_1, \delta_2)$. The BPS limit imposes a condition
\be\label{eq:AdS4BPSCond}
  e^{2 \delta_1 + 2 \delta_2} = 1 + \frac{2}{a g}\, .
\ee
For the black hole solution to have a regular horizon, we impose an additional constraint
\be\label{eq:AdS4RegCond}
  (m g)^2 = \frac{\textrm{cosh}^2 (\delta_1 + \delta_2)}{e^{\delta_1 + \delta_2}\, \textrm{sinh}^3 (\delta_1 + \delta_2)\, \textrm{sinh} (2 \delta_1)\, \textrm{sinh} (2 \delta_2)}\, .
\ee
The two conditions \eqref{eq:AdS4BPSCond} \eqref{eq:AdS4RegCond} in \cite{Cvetic:2005zi} have typos, which have been corrected in \cite{Chow:2013gba, Choi:2018fdc}, see also \cite{Cassani:2019mms}. With these constraints, there are only two independent parameters for asymptotically AdS$_4$ BPS black holes, which we choose to be $(\delta_1, \delta_2)$ for convenience.  In the BPS limit, the position of the outer horizon is
\be\label{eq:AdS4 r+}
  r_+ = \frac{2 m\, \textrm{sinh} (\delta_1)\, \textrm{sinh} (\delta_2)}{\textrm{cosh} (\delta_1 + \delta_2)}\, ,
\ee
which coincides with the inner horizon.

The physical quantities of non-extremal  AdS$_4$ black holes can also be solved as functions of $(a, m, \delta_1, \delta_2)$. In particular, the gravitational angular velocity $\Omega$ and the temperature $T$ are given by
\be
  \Omega = \frac{a (1 + g^2 r_1 r_2)}{r_1 r_2 + a^2}\, ,\quad T = \frac{\Delta'_r}{4 \pi (r_1 r_2 + a^2)}\, .
  \label{eq:Temperature of BPS}
\ee
Moreover, the other thermodynamic quantities of asymptotically AdS$_4$ black holes are \cite{Cvetic:2005zi}
\begin{align}
\begin{split}\label{eq:AdS4thermo}
  S & = \frac{\pi (r_1 r_2 + a^2)}{\Xi}\, ,\\
  J & = \frac{m a}{2 \Xi^2} \left(\textrm{cosh} (2 \delta_1) + \textrm{cosh} (2 \delta_2) \right)\, ,\\
  Q_1 = Q_2 & = \frac{m}{4 \Xi}\, \textrm{sinh} (2 \delta_1)\, ,\\
  Q_3 = Q_4 & = \frac{m}{4 \Xi}\, \textrm{sinh} (2 \delta_2)\, .
\end{split}
\end{align}

The 4d $\mathcal{N}=4$ gauged supergravity can be obtained by the truncation of the 11d supergravity, it admits supersymmetric black holes with two scalar and two Maxwell fields \cite{Chong:2004na}. A microscopic foundation for the entropy of these black holes was provided in 
\cite{Choi:2019zpz,Nian:2019pxj} via the AdS/CFT correspondence. Inspired by the successful studies of logarithmic in area corrections to the Bekenstein-Hawking entropy of asymptotically flat black holes pioneered by Sen (a different one-loop contribution $\sim \log (S_0)$ \cite{Sen:2012kpz,Sen:2012cj,Sen:2012dw}, the question of such logarithmic corrections for AdS black holes was addressed in \cite{Liu:2017vll, Jeon:2017aif,Liu:2017vbl,PandoZayas:2020iqr}. In this case the treatment of zero modes is more subtle that in asymptotically flat black holes. In the spherically symmetric case there are two types of zero modes. There  is a two-form zero mode in AdS$_4$, however, in the AdS$_2\times S^2$ near-horizon geometry there are ``other zero modes'' (graviton, one-form, gravitino). Performing the one-loop computation in the near-horizon geometry  \cite{Liu:2017vll, Jeon:2017aif} does not agree with the field theory answer. This circumstance has been crucial in successfully matching the logarithmic corrections to the entropy of AdS$_4\times S^7$ black holes to computations on the field theory side (ABJM) \cite{Liu:2017vbl} and various other AdS$_4$/CFT$_3$ pairs \cite{Gang:2019uay,Benini:2019dyp,PandoZayas:2020iqr}.

A systematic, bottom-up treatment of logarithmic area corrections  ($\sim \log S_0$) to AdS$_4$ black holes was presented in \cite{David:2021eoq} including for minimal gauged supergravity. An important conclusion relevant for extremal black holes was to demonstrate that the {\it local}  contribution to the one-loop effective action is the same whether it is computed from the full geometry or from the near-horizon geometry; only the contribution of zero modes is different from the full geometry and the near-horizon region. We hope to revisit the interplay of these $\log(S_0)$ versus $\log( T_{\rm Hawking})$ in the supersymmetric setups elsewhere. 

%%%%%%%%%%%%%%%%%%%%%%%%%%%%%%%%%%%%%%%%%%
\subsection{Ensemble choice and near-horizon limit}
To obtain the near-horizon and low-temperature geometry, we first choose an ensemble. As we expect the one-loop correction to the low-temperature entropy does not depend on the ensemble choice, we choose the ensemble for calculation convenience. In this ensemble, the relation \eqref{eq:AdS4RegCond} is always kept. The parameter $a$ and the radius of the outer horizon $r_+$ depend on the temperature,
\bea
a(T)=\frac{2}{g(e^{2 \delta_1 + 2 \delta_2}-1)}+c_1 T+c_2 T^2+c_3 T^3+O(T^4),\\
r_+(T)=\frac{2 m\, \textrm{sinh} (\delta_1)\, \textrm{sinh} (\delta_2)}{\textrm{cosh} (\delta_1 + \delta_2)}+d_1 T+d_2 T^2+d_3 T^3+O(T^4),
\eea
where the coefficients $c_i,d_i$ can be obtained by solving the second equation of \eqref{eq:Temperature of BPS} and $\Delta_{r}=0$. 
We verify that $c_1=0$. Other coefficients are lengthy and not illuminating to be written here.

To zoom into the near-horizon region, we conduct the following coordinate transformation:
\be
r=r_+(T)+T k_1(y-1),\ t=\frac{-i\tau}{k_2 T},\ \phi=\varphi+\frac{-ik_3\tau}{T}+k_4i\tau,
\ee
where the precise expressions for the constants $k_i$ are as follows:

\begin{align}
k_2 &= 2 \pi, \\
k_3 &= \frac{(-3 + e^{2 \delta_1 + 2 \delta_2}) g}{(-1 + e^{2 \delta_1 + 2 \delta_2}) k_2}, \\
\notag
k_1 &= (4 e^{2 (\delta_1 + \delta_2)} (-1 + e^{4 \delta_1}) (-1 + e^{4 \delta_2}) (1 + e^{2 (\delta_1 + \delta_2)}) \pi)/(-2 e^{4 \delta_1} - 2 e^{4 \delta_2}\\
\notag
&- 3 e^{2 (\delta_1 + \delta_2)} + 14 e^{4 (\delta_1 + \delta_2)} - 10 e^{6 (\delta_1 + \delta_2)} + 6 e^{8 (\delta_1 + \delta_2)} + e^{10 (\delta_1 + \delta_2)} + 5 e^{6 \delta_1 + 2 \delta_2}\\
& - 8 e^{8 \delta_1 + 4 \delta_2} + 5 e^{2 \delta_1 + 6 \delta_2} + e^{10 \delta_1 + 6 \delta_2} - 8 e^{4 \delta_1 + 8 \delta_2} + e^{6 \delta_1 + 10 \delta_2} g^2), \\
\notag
k_4 &=(-1 + e^{4 \delta_1}) (e^{2 \delta_1} + e^{2 \delta_2}) (-1 + e^{4 \delta_2}) (-3 + e^{2 (\delta_1 + \delta_2)}) (-1 + e^{2 (\delta_1 + \delta_2)})^3 \\&
\sqrt{e^{-\delta_1 - \delta_2} \coth^2(\delta_1 + \delta_2) \csch(2 \delta_1) \csch(2 \delta_2) \csch(\delta_1 + \delta_2)}/(2 (1 + e^{2 (\delta_1 + \delta_2)})\\ 
\notag
&(-2 e^{4 \delta_1} - 2 e^{4 \delta_2} - 3 e^{2 (\delta_1 + \delta_2)} + 14 e^{4 (\delta_1 + \delta_2)} - 10 e^{6 (\delta_1 + \delta_2)} + 6 e^{8 (\delta_1 + \delta_2)} + e^{10 (\delta_1 + \delta_2)}\\
& + 5 e^{6 \delta_1 + 2 \delta_2} - 8 e^{8 \delta_1 + 4 \delta_2} + 5 e^{2 \delta_1 + 6 \delta_2} + e^{10 \delta_1 + 6 \delta_2} - 8 e^{4 \delta_1 + 8 \delta_2} + e^{6 \delta_1 + 10 \delta_2})).
\end{align}
By taking the $T\to0$ limit, we obtain the near-horizon geometry on the zeroth order of~$T$:
\be
\label{eq:BPS-AdS4 NearHorizon Zeroth}
\Bar{g}_{\mu\nu}dx^\mu dx^\nu =  g_1(\theta ) \left(\frac{dy^2}{y^2-1} + d\tau ^2 \left(y^2-1\right)\right)+   g_2( \theta)d\theta^2 +g_3( \theta) \Big(d \varphi+i k(y-1)  d\tau   \Big)^2,
\ee
where
\be
\begin{aligned}
    g_1(\theta)=&\frac{e^{2 (\delta_1 + \delta_2)} + \cos(2 \theta)}{2 \left(1 + e^{2 (\delta_1 + \delta_2)}\right) \pi}k_1,\\
    g_2(\theta)=&- \frac{2 \left(e^{2 (\delta_1 + \delta_2)} + \cos(2 \theta)\right)}{g^2 \left(1 + 3 e^{2 (\delta_1 + \delta_2)} - 3 e^{4 (\delta_1 + \delta_2)} + e^{6 (\delta_1 + \delta_2)} + 2 \cos(2 \theta)\right)},\\
    g_3(\theta)=&\frac{2 \left(-1 - 2 e^{2 (\delta_1 + \delta_2)} + e^{4 (\delta_1 + \delta_2)} - 2 \cos(2 \theta)\right) \sin^2(\theta)}{\left(-3 + e^{2 (\delta_1 + \delta_2)}\right)^2 g^2 \left(e^{2 (\delta_1 + \delta_2)} + \cos(2 \theta)\right)},\\
    k=&-k_4.& 
\end{aligned}
\ee
By keeping $O(T)$ when $T\to0$, we can also obtain the near-horizon geometry on the first order of temperature $\delta g_{\mu\nu}$. Again, this expression is lengthy and not illuminating to be provided here. 
%%%%%%%%%%%%%%%%%%%%%%%%%%%%%%%%%%%%%%%%%%%%%%
\subsection{Lichnerowicz operator and the zero modes}
The Lichnerowicz operator of our action \eqref{eq:sugra action} is
\bea
\notag
h_{\alpha\beta}\Delta^{\alpha \beta, \mu \nu}_{L}h_{\mu\nu} &=& h_{\alpha \beta}(\Delta_{EH}^{\alpha\beta,\mu\nu}-\frac{1}{2} e^{2 \zeta}\Delta_{F_2}^{\alpha\beta,\mu\nu}-\frac{e^{\zeta}}{2\left(1+\chi^{2} e^{2 \zeta}\right)}\Delta_{F_1}^{\alpha\beta,\mu\nu}-\frac{1}{2}\Delta_{\zeta}^{\alpha\beta,\mu\nu}-\frac{1}{2}e^{2\zeta}\Delta_{\chi}^{\alpha\beta,\mu\nu}\\
&+&g^2(4+2\cosh{\zeta}+e^{\zeta}\chi^2)\Delta_{1}^{\alpha\beta,\mu\nu}) h_{\mu \nu},
\label{eq:BPS-AdS Linch operator}
\eea
where the different contributions to the Lichnerowicz operator can be found in \ref{appendix:Lich operator}.

At zero temperature, i.e. when $g=\bar{g}$, one can verify that the above operator has the following zero modes:
\be
h_{\mu\nu}^{(n)}dx^\mu dx^\nu=c_n e^{i n \tau} \left(\frac{-1 + y}{1 + y}\right)^{\frac{\abs{n}}{2}} \left(e^{2(\delta_1+\delta_2)}+ \cos(2 \theta)\right)(d\tau^2-2i\frac{\abs{n}}{n}\frac{d\tau dy}{y^2-1}-\frac{dy^2}{(y^2-1)^2})
\label{eq:BPS-AdS zero modes}
\ee
for $\abs{n}\geq2 $, where $c_n$ is the normalization constant.

%%%%%%%%%%%%%%%%%%%%%%%%%%%%%%%%%%%%%%%%
%\subsection{The lifted eigenvalue and entropy correction}
By turning on a small temperature, the ``zero modes'' will be lifted and gain a nonzero eigenvalue. Now we substitute $g=\bar{g}+\delta g$ into the Lichnerowicz operator \eqref{eq:BPS-AdS Linch operator} and apply it to the zero modes \eqref{eq:BPS-AdS zero modes}. The lifted eigenvalue is
\be
\delta\lambda_n=\int dx^4 \sqrt{\bar{g}}h^{(n)}_{\alpha\beta}\delta\Delta_L^{\alpha\beta,\mu\nu} h^{(n)}_{\mu\nu}.
\ee
The expression of the operator is intractable, but the integration is straight forward and the result is simple:
\be \ \label{eq:correctedeigenval_KerrAdSNewmanGaugedSUGRA}
\delta\lambda_n\propto nT,\ \abs{n}\geq2.
\ee
Using Zeta function regularization, the contribution of the extremal zero modes to the low-temperature partition function is therefore:
\be
\delta \log Z=\log(\prod_{n\geq2}\frac{\pi}{\delta\lambda_n})=\frac{3}{2}\log(\frac{T}{T_q})+O(1),
\ee
where the $O(1)$ term is not reliable. The temperature scale $T_q$ can be extracted from the $O(T)$ coefficient of classical entropy, its precise form is
\bea
\notag
T_q=&\bigg(&2 \left(-3 + e^{2 (\delta_1 + \delta_2)}\right) \left(1 + e^{2 (\delta_1 + \delta_2)}\right) (-2 e^{4 \delta_1} - 2 e^{4 \delta_2} - 3 e^{2 (\delta_1 + \delta_2)} + 14 e^{4 (\delta_1+ \delta_2)}\\ \notag
&-& 10 e^{6 (\delta_1 + \delta_2)} + 6 e^{8 (\delta_1 + \delta_2)} + e^{10 (\delta_1 + \delta_2)} + 5 e^{6 \delta_1 + 2 \delta_2}- 8 e^{8 \delta_1 + 4 \delta_2} + 5 e^{2 \delta_1 + 6 \delta_2}\\ \notag
&+& e^{10 \delta_1 + 6 \delta_2}- 8 e^{4 \delta_1+ 8 \delta_2} + e^{6 \delta_1 + 10 \delta_2}) g^3 \bigg)/\bigg(4\pi^2\left(-1 + e^{4 \delta_1}\right) \left(e^{2 \delta_1} + e^{2 \delta_2}\right) \left(-1 + e^{4 \delta_2}\right) \\
&(&-1 + e^{2 (\delta_1 + \delta_2)})^4 \sqrt{e^{-\delta_1 - \delta_2} \coth^2{\left(\delta_1 + \delta_2\right)} \csch{2 \delta_1} \csch{2 \delta_2} \csch{\left(\delta_1 + \delta_2\right)}}\bigg).
\eea
This quantity is extracted from the  classical heat capacity and  agrees (in the $\delta_1=\delta_2=\delta\neq0$ limit) with the expression presented in \cite{David:2020jhp,Larsen:2020lhg}.

It is worth emphasizing that our investigation in this paper focuses strictly in the contribution of the graviton zero modes to the path integral. A {\it bona fide} discussion of the black hole in the supersymmetric BPS limit, requires a proper treatment of the gravitino and other zero modes which is expected to changed the resulting logarithmic in temperature contribution. Here we want to highlight {\it only} the universality of the graviton contribution and that is the main purpose that this section serves. A systematic treatment, including the effects of fermionic and vector zero modes will be presented elsewhere.
%%%%%%%%%%%%%%%%%%%%%%%%%%%%%%%%%%%%%%%%%%%%%%%%
\section{Myers-Perry black holes in five dimensions} \label{Sec:MyersPerryBlack hole}

In this section, we study a vacuum solution of the Einstein's equations that describes spinning black holes, known as the Myers-Perry (MP) black holes. Some aspects of the connection of five-dimensional rotating black holes with JT gravity were discussed in \cite{Castro:2018ffi,Moitra:2019bub} with emphasis on classical aspects. Our goal in this manuscript is to focus on the quantum contribution to the low-temperature thermodynamics. 

The MP black hole \cite{MYERS1986304, Myers:2011yc} in general odd $d~ (\geq5)$ dimensions with $d=2n+1$ is described by the metric
\begin{eqnarray} \label{eq:MyersPerryBH}
    ds^2 = -dt^2 + \frac{\mu r^2}{\Pi F} \left(dt + \sum^n_{i=1}a_i \mu^2_i d\phi_i\right)^2 + \frac{\Pi F}{\Pi - \mu r^2}dr^2 + \sum^n_{i=1} \left(r^2+a_i^2\right)\left(d\mu^2_i +\mu^2_i d\phi^2_i\right)\,, \qquad
\end{eqnarray} where the parameters in the metric are given by
\begin{eqnarray}\label{eq:MyersPerryBHparameters}
    F=1- \sum_{i=1}^n\frac{a^2_i \mu^2_i}{r^2+a_i^2}\, ,\qquad  \Pi = \prod_{i=1}^n \left(r^2 +a_i^2\right)\, ,
\end{eqnarray} where $\mu$ is a free parameter in the theory proportional to the mass of the black hole, $a_i$'s are related to the angular velocity of the rotating black hole and $\{\mu_i\}$ satisfy the constraint $\sum^n_{i=1} \mu_i^2 = 1$. The conserved charges of this class of black holes are given by
\begin{eqnarray}
    M=\frac{\left(d-2\right) \Omega_{d-2}}{16 \pi} \mu \, , \qquad J^{y_i x_i} = \frac{2}{d-2} M a_i\, ,
\end{eqnarray} where $\Omega_{d}$ is the volume of the $d$ sphere. Naturally, setting $a_i$ to zero takes us to the Schwarzschild metric whereas setting $M$ to zero gives us the flat space vacuum limit.

In this section, we focus only on the $d=5$ case, and the metric is obtained by setting $n=2$ in equations \eqref{eq:MyersPerryBH} and \eqref{eq:MyersPerryBHparameters}. Setting $a_1 = a, ~a_2= b$ and parametrizing  $\mu_i$ as $\mu_1 = \sin \theta\, , ~ \mu_2 = \cos \theta $, we obtain the metric as 
\begin{eqnarray} \label{eq:MP5metric}
    ds_{(5d
    )}^2 &= -dt^2 + \frac{\mu}{\Sigma} \left(dt+ a  \sin^2 \theta d\phi_1 + b  \cos^2 \theta d\phi_2   \right)^2 + \left(\frac{\Sigma }{\Pi - \mu r^2}\right) r^2 dr^2 + \, \nonumber \\
   & \Sigma d\theta^2 + \left(r^2 + a^2\right) \sin^2 \theta d\phi_1^2 + \left(r^2 + b^2\right) \cos^2 \theta d\phi_2^2 \,,   
\end{eqnarray} where 
\begin{eqnarray}
    \Sigma = r^2 + a^2 \cos^2 \theta + b^2 \sin^2 \theta \, , \qquad \Pi = \left(r^2 +a^2\right)\left(r^2 +b^2\right)\, . 
\end{eqnarray}

To calculate $\log T $ corrections, as before, we need to move slightly away from extremality, thereby leading to a perturbative series for the metric, with the perturbing parameter being the temperature $T$ associated with the outer horizon. Below, we proceed to calculate the near-horizon limit of $5d$ Myers-Perry black hole.

\paragraph{Equal angular momenta:} For ease of computation, we first focus on the case of equal angular momenta, i.e., we set $a=b$. In the following we use $2\pi T$ in place of the temperature. We see that the horizon is obtained as a root of the equation
\begin{eqnarray}\label{eq:horizonisrootof}
    \Pi - \mu r^2=0\, 
\end{eqnarray} and this immediately gives us the inner ($r_{-}(T)$) and outer horizon ($r_{+}(T)$) as 
\begin{eqnarray} \label{eq:innerhorizon}
   r_{-}(T)=\frac{1}{\sqrt{2}}\sqrt{\mu -\sqrt{\mu  \left(\mu -4 a^2\right)}-2 a^2 }\,, \\ \label{eq:outerhorizon}
   r_{+}(T) = \frac{1}{\sqrt{2}}\sqrt{\mu +\sqrt{\mu  \left(\mu -4 a^2\right)}-2 a^2 } \, . 
\end{eqnarray} The temperature itself is given by
\begin{eqnarray}
    T = \left(\frac{\partial_r \Pi - 2 \mu r}{2 \mu r^2}\right)\Bigg|_{r=r_{+}} = \frac{ \sqrt{2\mu  \left(\mu -4 a^2\right)}}{\mu  \sqrt{\sqrt{\mu  \left(\mu -4 a^2\right)}-2 a^2+\mu }}\, .
\end{eqnarray} This induces a $T$ correction to the mass and angular velocity, and they are given by 
\begin{eqnarray}
    \mu &=& \frac{4 J^{2/3}}{\pi ^{2/3}} + \frac{4 J^{4/3} T^2}{3 \pi ^{4/3}}+ \frac{8 J^{5/3} T^3}{3 \pi ^{5/3}}+  \ldots\, , \\
    a&=& \frac{ J^{1/3}}{\pi ^{1/3}}-\frac{J T^2}{3 \pi }- \frac{2 J^{4/3} T^3}{3 \pi ^{4/3}}  + \ldots\, .
\end{eqnarray} Note that the corrections are such that
\begin{eqnarray}
    J=\frac{\Omega_3}{8\pi} \mu a 
\end{eqnarray} is actually kept fixed at its extremal value, i.e., it receives no $T$ corrections. This ensures that we are manifestly working in the canonical ensemble where the angular momenta are kept fixed. Substituting back in equations \eqref{eq:innerhorizon} and \eqref{eq:outerhorizon}, we obtain the following series for the inner and outer horizon radii
\begin{eqnarray}
    r_{-}(T) = \left(\frac{J}{\pi}\right)^{\frac{1}{3}} - \left(\frac{J}{\pi}\right)^{\frac{2}{3}} T - \frac{5J}{6\pi} T^2 + \ldots\, \, , \\
     r_{+}(T) = \left(\frac{J}{\pi}\right)^{\frac{1}{3}} + \left(\frac{J}{\pi}\right)^{\frac{2}{3}} T + \frac{7J}{6\pi} T^2 + \ldots\, \, . 
\end{eqnarray}
%%%%%%%%%%%%%%%%%%%%%%%%%%%%%%%%%%%%%%%%%%%%%%%%%%%%%
\subsection{Near horizon limit of the MP black hole}
In this section we do a coordinate transformation to probe the near horizon geometry of the MP black hole when it is slightly out of extremality.

Following \cite{Galajinsky:2012vh,Galajinsky:2013mla}, we probe the near horizon geometry of 5$d$ MP black hole as follows. We do a coordinate transformation given by
\begin{equation}
   t\rightarrow  -\frac{i\tau}{T}, \quad r \rightarrow r_{+}\left(T\right) + r_0^2\, T \left(y-1\right), \quad \phi_{1,2} \rightarrow \varphi_{1,2} + i \left(\frac{1}{2 r_0 T} - 1 \right) \tau.
\end{equation} 
Here, $r_0$ is the radius of the outer horizon at extremality which is related to the angular momentum as
\begin{equation}
    r_0 = \left(\frac{J}{\pi}\right)^{\frac{1}{3}}.
\end{equation}
This transformation yields the near horizon metric to be 
\begin{eqnarray} \label{eq:NHEK Myers Perry}
O\left(1\right):\quad  ds_{(0)}^2 = && \bar{g}_{\mu\nu}dx^\mu dx^\nu = \frac{J^{\frac{2}{3}}}{2\pi^{\frac{2}{3}}} \left( \frac{dy^2}{y^2-1} + \left(y^2-1\right)d\tau^2 \right) + \frac{2J^{\frac{2}{3}}}{\pi^{\frac{2}{3}}} d\theta^2 \nonumber \\
&& + \frac{2 J^{2/3} \sin^2\theta  ((y-1) \cos 2 \theta + 3 y-5)}{\pi ^{2/3} (y-2)} \left( d_1 + \frac{i\left(y-2\right)}{2}d\tau \right)^2 \nonumber \\ && + \frac{2 J^{2/3} \cos ^2\theta \left(\left(y-1\right) \cos2\theta + 5y - 7 \right)}{\pi^{2/3} \left( 3y - 4 \right)} \left( d\varphi_2 + \frac{i\left(y-2\right)}{2}d\tau \right)^2 \nonumber \\ && - \frac{2 J^{2/3} (y-2) \sin^2\theta \cos^2\theta}{\pi^{2/3} (3y-4)} \left( i\left(y-1\right)d\tau^2 + \frac{3y-4}{y-2}d\varphi_1^2 + d\varphi_2^2 \right). \qquad
\end{eqnarray} We also obtain the corresponding correction to the near-horizon extreme Myers-Perry (NHEMP) metric given by
\begin{align}
  O\left(T\right): ds_{(1)}^2 = T g^{(1)}_{\mu\nu} dx^\mu dx^\nu\,,
\end{align} where, as in section \ref{Sec:Kerr-AdS}, for the sake of presentation, we do not explicitly type the corrections here. With the corrected metric in hand, we now proceed to evaluate the zero modes of the Lichnerowicz operator.

\subsection{Zero modes and correction to black hole entropy} We begin by choosing an ansatz for the diffeomorphism generating vector field given by
\begin{equation}
    \xi^{(n)} = e^{i n \tau} \left(f_2\left(y\right)\partial_{\tau} + f_1\left(y\right)\partial_y + f_3\left(y\right)\partial_{\varphi_1} + f_4\left(y\right)\partial_{\varphi_2} \right). 
\end{equation} 
  With this ansatz, we construct the zero modes of the Lichnerowicz operator given by~$h_{\mu\nu} = \mathcal{L}_{\xi} \bar g_{\mu\nu}$.   We further demand that the $h_{\mu\nu}$ so generated belong to the kernel of the Lichnerowicz operator. This fixes the functions $f_i (y)$ to be 
\begin{align}
    f_1\left(y\right) &= \left(\frac{y-1}{y+1}\right)^{\frac{|n|}{2}} \frac{|n|+y}{2 \left(n^2-1\right)}\,,\\
    f_2\left(y\right) &= i \left(\frac{y-1}{y+1}\right)^{\frac{|n|-2}{2}} \frac{2+|n|+n^2\left(y-2\right)-2 |n| y - 2 y^2}{4 |n| \left(n^2-1\right) \left(y+1\right)^2},\\
    f_3\left(y\right) &= \left(\frac{y-1}{y+1}\right)^{\frac{|n|-2}{2}} \frac{2+|n|+n^2\left(y-2\right)-2 |n| y - 2 y^2}{4 |n| \left(n^2-1\right) \left(y+1\right)^2}\,,\\
        f_4\left(y\right) &= \left(\frac{y-1}{y+1}\right)^{\frac{|n|-2}{2}} \frac{2+|n|+n^2\left(y-2\right)-2 |n| y - 2 y^2}{4 |n| \left(n^2-1\right) \left(y+1\right)^2}= f_3\left(y\right)\,,
\end{align} and
\begin{equation} \label{eq:MP5eqzeromodes}
    h_{\mu\nu}^{(n)}dx^\mu dx^\nu = \frac{\sqrt{|n| \left(n^2-1\right)}e^{i n \tau}}{8 \pi ^{4/3} J^{1/6}} \left(\frac{y-1}{y + 1}\right)^{\frac{|n|}{2}} \left(-d\tau^2+2i\frac{|n|}{n}\frac{d\tau dy}{y^2-1}+\frac{dy^2}{(y^2-1)^2} \right) \,,
\end{equation} for $|n| \geq 2$. Hereafter, the steps to be performed mimic section \ref{sec:logTAdS} and the corrected eigenvalue of the Lichnerowicz operator at first order is given by
    \begin{equation}
        \delta\lambda_n = \int d^4x \sqrt{\bar{g}}\: \left(h^{(n)}\right)_{\alpha\beta} \left(\delta \Delta^{\Lambda}\right)^{\alpha\beta, \mu\nu} \left(h^{(n)}\right)_{\mu\nu}\, . 
\label{eq:LiftedEigenvalueMyersPerry}
    \end{equation}
    Explicit computation  of \eqref{eq:LiftedEigenvalueMyersPerry} gives us 
\begin{equation} \label{eq: LiftedEigenvalueMyersPerry_ II}
    \delta\lambda_n = \frac{n\pi^{4/3}T}{8 J^{1/3}},
\end{equation} 
where we reverted back to the actual temperature $T$ of the black hole by appropriately inserting factors of $2\pi$ (see statement before equation \eqref{eq:horizonisrootof}).

Thereafter, following arguments similar to the one in section \ref{sec:logTAdS} and the Zeta function regularization, we have 
\begin{eqnarray}
    \log Z \Bigg|_{\log T} = -\log \Bigg[\prod^\infty_{n=2} \left(\frac{n \pi^{\frac{4}{3}}  T}{8J^{\frac{1}{3}}}\right) \Bigg] \sim \frac{3}{2} \log T \, , \qquad 
\end{eqnarray} where once again we have ignored irrelevant constants. 

%%%%%%%%%%%%%%%%%%%%%%%%%%%%%%%%%%%%%%%%%%%%%%%
\subsection{Comment on the case of two different angular momenta} Here, we summarize a few statements about the $5d$ Myers Perry black with different angular momenta $\left(J_1 \neq J_2\right)$. The calculation proceeds exactly as we describe in the previous section with added computational complexity. We find that the outer horizon now admits the small-temperature expansion
\begin{equation}
    r_{+}\left(T\right) = \frac{2^{2/3} \sqrt{J_1 J_2}}{\pi^{1/3}(J_1+J_2)^{2/3}} + \frac{(J_1+J_2)^{2/3}}{(2 \pi )^{2/3}} T + \frac{\left(9 J_1^2+10 J_1 J_2+9 J_2^2\right) \sqrt{J_1 J_2}}{24 \pi  J_1 J_2} T^2 + \ldots\, . 
\end{equation}
As before, we start from the metric \eqref{eq:MP5metric}, and do the following coordinate transformations
\begin{equation}
    t \to -\frac{i\tau}{T}, \quad r\to r_{+}\left(T\right) + r_0^2\, T \left(y-1\right), \quad \phi_{1,2} \to \varphi_{1,2} + \frac{2\sqrt{J_1 J_2}}{J_1+J_2}\left(\frac{1}{2r_0T}-1\right)i\tau,
\end{equation}
to zoom into the near-horizon region. Upto linear order in temperature, we obtain the near horizon extreme geometry
\begin{equation}
\begin{split} 
\label{eq:nearhorizonextremeMP}
    ds_{(0)}^2 =\bar g_{\mu\nu} dx^\mu dx^\nu =&\bar{g}_{11}\left(y,\theta\right) d\tau^2 + 2 \bar{g}_{14}\left(y,\theta\right)\, d\tau d\phi_1 + 2 \bar{g}_{15}\left(y,\theta\right)\, d\tau d\phi_2 + \bar{g}_{22}\left(y,\theta\right) dr^2 \\ &+ \bar{g}_{33}\left(\theta\right) d\theta_1^2 + \bar{g}_{44}\left(\theta\right)d\phi_1^2 + 2 \bar{g}_{45}\left(\theta\right)d\phi_1 d\phi_2\,,
\end{split}
\end{equation}
and its first-order temperature correction (which we do not write explicitly)
\begin{equation}
    ds_{(1)}^2 = T\,g_{\mu\nu}^{(1)}dx^{\mu}dx^{\nu}.
\end{equation}
Here the functions entering the near-horizon extreme metric \eqref{eq:nearhorizonextremeMP} are given by
\begin{equation}
\begin{split}
    \bar g_{11}\left(y,\theta \right) = -&\Big(\left(J_1+J_2 + \left(J_1-J_2\right)\cos{2\theta} \right)^2 \left(3 J_1^6+22 J_1^5 J_2-192 J_1^4 J_2^2 y+253 J_1^4 J_2^2 + \right. \\ &\left. 128 J_1^3 J_2^3 y^2-640 J_1^3 J_2^3 y+596 J_1^3 J_2^3-192 J_1^2 J_2^4 y+253 J_1^2 J_2^4+ \right. \\ &\left. 4 \cos{2\theta} \left(J_1+J_2\right)^3 \left(J_1^3-15 J_1^2 J_2+15 J_1 J_2^2-J_2^3\right) + 22 J_1 J_2^5 + \right. \\ &\left.\cos{4\theta} \left(J_1-J_2\right)^2 \left(J_1+J_2\right)^4+3 J_2^6\right)\Big) \Big(32 \left(2 \pi \right)^{2/3} \left(J_1+J_2\right)^{4/3} \times \\ &\left(J_1^2 \cos^2{\theta} + J_2 \left(J_1+J_2 \sin^2{\theta} \right)\right)^3\Big)^{-1},
\end{split}
\end{equation}
\begin{equation}
\begin{split}
    \bar g_{14}\left(y,\theta\right) = &2^{1/3} i \sin^2{\theta} \sqrt{J_1 J_2} \Big(-5 J_1^3+\cos{2\theta}\left(J_1-J_2\right) \left(J_1^2+2 J_1 J_2 \left(2 y-1\right)+J_2^2\right) + \\ &3 J_1^2 J_2 \left(4 y-7\right)+J_1 J_2^2 \left(4 y-7\right)+J_2^3\Big) \Big(\pi ^{2/3} \left(J_1+J_2\right)^{7/3} \times \\ &\left(\cos{2\theta} \left(J_1-J_2\right)+J_1+J_2\right)\Big)^{-1},
\end{split}
\end{equation}
\begin{equation}
\begin{split}
    \bar g_{15}\left(y,\theta\right) = &2^{1/3} i \sin^2{\theta} \sqrt{J_1 J_2} \Big(-5 J_2^3+\cos{2\theta}\left(J_1-J_2\right) \left(J_1^2+2 J_1 J_2 \left(2 y-1\right)+J_2^2\right) + \\ &3 J_2^2 J_1 \left(4 y-7\right)+J_2 J_1^2 \left(4 y-7\right)+J_1^3\Big) \Big(\pi ^{2/3} \left(J_1+J_2\right)^{7/3} \times \\ &\left(\cos{2\theta} \left(J_1-J_2\right)+J_1+J_2\right)\Big)^{-1},
\end{split}
\end{equation}
\begin{align}
    \bar{g}_{22}\left(y,\theta\right) &= \frac{J_1 J_2 (J_1+J_2+\left(J_1-J_2\right)\cos{2\theta})}{(2 \pi )^{2/3} (y-1) \left(J_1+J_2\right)^{1/3} \left(J_1^2+2 y J_1 J_2 + J_2^2\right)},\hspace{3.8cm}\\
    \bar{g}_{33}\left(\theta\right) &= \frac{2^{1/3}\left(J_1+J_2+(J_1-J_2)\cos{2\theta}\right)}{\pi ^{2/3} \left(J_1+J_2\right)^{1/3}},\\
  \bar{g}_{44}\left(\theta\right) &= \frac{2^{1/3} J_1 \sin ^2{\theta} (J_1+J_2)^{2/3} \left(2 J_1-J_2 \cos{2 \theta}+J_2\right)}{\pi ^{2/3} \left(J_1^2 \cos ^2(\theta )+J_2 \left(J_1+J_2 \sin ^2\theta\right)\right)},\\
    \bar{g}_{55}\left(\theta\right) &= \frac{2^{1/3} J_2 \cos ^2(\theta ) (J_1+J_2)^{2/3} \left(J_1 \cos{2 \theta} + J_1 + 2 J_2 \right)}{\pi^{2/3} \left(J_1^2 \cos^2{\theta} + J_2 \left(J_1+J_2 \sin^2\theta \right)\right)},\\
   \bar{g}_{45}\left(\theta\right) &= \frac{J_1 J_2 \sin^2{2 \theta} (J_1+J_2)^{2/3}}{(2 \pi )^{2/3} \left(J_1^2 \cos^2{\theta} + J_2 \left(J_1+J_2 \sin^2{\theta}\right)\right)}\,.
\end{align}
The diffeomorphism generating vector fields are given by
\begin{equation}
    \xi = e^{in\tau}\left(f_2\left(y\right)\partial_{\tau} + f_1\left(y\right)\partial_{y} + f_3\left(y\right)\partial_{\varphi_1} + f_4\left(y\right)\partial_{\varphi_2}\right),
\end{equation} where
\begin{eqnarray} \label{eq:f1ingenMyersPerry}
    f_1 &=& \frac{
    1}{2 \left(n^2-1\right)}\left(\frac{\Theta(J_1,J_2;y)-1}{\Theta(J_1,J_2;y)+1}\right)^{\frac{|n|}{2}}  
    \left(|n|+ \Theta(J_1,J_2;y)\right) 
     \,, \\
   f_2&=&\frac{i \left(f_1'(y) \bar g_{15}^{(1,0)}(y,\theta )+f_1(y) \bar g_{15}^{(2,0)}(y,\theta )\right)}{|n|\, \bar g_{15}^{(1,0)}(y,\theta )}\, , \\
    f_3 &=& \frac{i \left(f_1 \bar g_{45}(\theta ) \bar g_{15}^{(1,0)}(y,\theta )-f_1 \bar g_{55}(\theta ) \bar g_{14}^{(1,0)}(y,\theta )\right)+|n|\, f_2 \bar g_{55}(\theta ) \bar g_{14}(y,\theta )-|n|\, f_2 \bar g_{45}(\theta ) \bar g_{15}(y,\theta )}{|n| \left(\bar g_{45}(\theta )^2-\bar g_{44}(\theta ) \bar g_{55}(\theta )\right)}\,, \nonumber \\ \\
    f_4 &=&\frac{i f_1 \bar g_{15}^{(1,0)}(y,\theta )-|n|\, f_2 \bar g_{15}(y,\theta )-|n|\,f_3 \bar g_{45}(\theta )}{|n|\, \bar g_{55}(\theta )}\,,
\end{eqnarray} with
\begin{eqnarray}
\Theta(J_1,J_2;y) = \frac{  \left(J_1^2+2 J_1 J_2 (2 y-1)+J_2^2\right)}{(J_1+J_2)^2 }\, .
\end{eqnarray} 
It can be checked that these diffeomorphism generating vectors are harmonic, i.e.,
\begin{eqnarray}
     \bar \Box ~\xi_ \mu = 0\,
\end{eqnarray} and, therefore, they satisfy the harmonic gauge fixing condition. The zero modes of the Lichnerowicz operator are given by
\begin{equation}
\begin{split}
 & h^{(n)}_{\mu\nu} = \frac{\sqrt{|n| \left(n^2-1\right)}}{8 (2 \pi^2 )^{2/3} (J_1 J_2)^{1/4}}\times \frac{2^{\frac{1}{3}+\frac{n}{2}} e^{i n \tau} \left( \frac{J_1 J_2 \left(y-1\right)}{J_1^2+J_2^2+2 y J_1 J_2} \right)^{\frac{|n|}{2}} \left(J_1+J_2+\left(J_1-J_2\right)\cos2\theta\right)}{\left(J_1+J_2\right)^{\frac{2}{3}}} ~\times ~\\ 
 &\left(-d\tau^2 + i\frac{|n|}{n}\frac{\left(J_1+J_2\right)^2\left(\frac{J_1 J_2 \left(y-1\right)}{J_1^2 + J_2^2 + 2 y J_1 J_2}\right)}{J_1 J_2 \left(y-1\right)^2} d\tau\,dy + \frac{\left(J_1+J_2\right)^4}{4\left(y-1\right)^2\left(J_1^2+J_2^2+2 y J_1 J_2\right)} dy^2\right)\, ,
\end{split}
\end{equation}
which are generated by $h_{\mu\nu} = \mathcal{L}_{\xi}\bar{g}_{\mu\nu}$, as usual. The zero modes also reduce to equation \eqref{eq:NHEK Myers Perry} when $J_{1,2} \rightarrow J$. The integration over these zero modes (as in equation \eqref{eqn: LiftedEigenvalue}) is quite intractable in this general case of different angular momenta. However, we see the $n$ dependence of $h_{\mu\nu}$ above is comparable with the equal angular momenta case. This similarity owes its roots to the similar functional form\footnote{Note that $\Theta(J,J;y)=y$.}~of the diffemorphism generating vector field in equation \eqref{eq:f1ingenMyersPerry} (to be compared with the corresponding one in equation \eqref{eq:MP5eqzeromodes}). Therefore, the corresponding log correction is anticipated to be
\begin{eqnarray}
    \log Z \Bigg|_{\log T}  \sim \frac{3}{2} \log T \, , \qquad 
\end{eqnarray} where $J_{1,2}$ simply alter the functional form of $J_{1,2}$ dependent constants.
\iffalse
As of now, we have checked that the diffeomorphism generated by the above vector field satisfies
\begin{eqnarray}
    \Delta h_{1,3}=\Delta h_{1,4}=\Delta h_{2,4}=\Delta h_{3,4}=\Delta h_{1,5}=\Delta h_{2,5}=\nonumber \\ 
    \Delta h_{3,5}=\Delta h_{4,5}= \Delta h_{3,3}=\Delta h_{4,4}=\Delta h_{5,5}=0\, . 
\end{eqnarray} 
\fi 
%%%%%%%%%%%%%%%%%%%%%%%%%%%%%%%%%%%%%%%%%%%%%%%%%%%%%%%%%
\section{Kerr-AdS$_5$ Black hole}
\label{sec:Kerr-AdS5}
In this section, we provide the story for the Kerr-AdS$_5$ black hole with equal angular momenta 
%\subsection{Set up}
We start from the Einstein-Hilbert action with cosmology constant in 5 dimensions:
\be
I=-\frac{1}{16\pi}\int d^5x\sqrt{g}(R-2\Lambda)+I_{boundary},
\ee
where $\Lambda=-6g^2$.
The Kerr-AdS$_5$ solution is given by
\be
\label{eq:KerrAdS5mentric1}
\begin{aligned}
d s^2= & -\frac{\Delta_\theta\left[\left(1+g^2 r^2\right) \rho^2 d t \right] d t}{\Xi_a \Xi_b \rho^2}+\frac{f}{\rho^4}\left(\frac{\Delta_\theta d t}{\Xi_a \Xi_b}-\omega\right)^2+\frac{\rho^2 d r^2}{\Delta_r}+\frac{\rho^2 d \theta^2}{\Delta_\theta} \\
& +\frac{r^2+a^2}{\Xi_a} \sin ^2 \theta d \phi_1^2+\frac{r^2+b^2}{\Xi_b} \cos ^2 \theta d \phi_2^2,
\end{aligned}
\ee
where
\be
\begin{aligned}
\nu & \equiv b \sin ^2 \theta d \phi_1+a \cos ^2 \theta d \phi_2, \\
\omega & \equiv a \sin ^2 \theta \frac{d \phi_1}{\Xi_a}+b \cos ^2 \theta \frac{d \phi_2}{\Xi_b}, \\
\Delta_\theta & \equiv 1-a^2 g^2 \cos ^2 \theta-b^2 g^2 \sin ^2 \theta, \\
\Delta_r & \equiv \frac{\left(r^2+a^2\right)\left(r^2+b^2\right)\left(1+g^2 r^2\right)}{r^2}-2 m, \\
\rho^2 & \equiv r^2+a^2 \cos ^2 \theta+b^2 \sin ^2 \theta, \\
\Xi_a & \equiv 1-a^2 g^2, \\
\Xi_b & \equiv 1-b^2 g^2, \\
f & \equiv 2 m \rho^2.
\end{aligned}
\ee
For the convenience of our following discussion, we rewrite the above solution \eqref{eq:KerrAdS5mentric1} in the following form \cite{David:2020ems}:
\be
d s^2=-\frac{\Delta_r \Delta_\theta r^2 \sin ^2(2 \theta)}{4 \Xi_a^2 \Xi_b^2 B_{\phi_1} B_{\phi_2}} d t^2+\rho^2\left(\frac{d r^2}{\Delta_r}+\frac{d \theta^2}{\Delta_\theta}\right)+B_{\phi_2}\left(d \phi_2+v_1 d \phi_1+v_2 d t\right)^2+B_{\phi_1}\left(d \phi_1+v_3 d t\right)^2,
\ee
where
\be
\begin{gathered}
B_{\phi_1} \equiv \frac{g_{33} g_{44}-g_{34}^2}{g_{44}}, \quad B_{\phi_2} \equiv g_{44}, \\
v_1 \equiv \frac{g_{34}}{g_{44}}, \quad v_2 \equiv \frac{g_{04}}{g_{44}}, \quad v_3 \equiv \frac{g_{04} g_{34}-g_{03} g_{44}}{g_{34}^2-g_{33} g_{44}},
\end{gathered}
\ee
with  the non-vanishing components of the metric \eqref{eq:KerrAdS5mentric1}
\be
\begin{aligned}
& g_{00}=-\frac{\Delta_\theta\left(1+g^2 r^2\right)}{\Xi_a \Xi_b}+\frac{\Delta_\theta^2\left(2 m \rho^2\right)}{\rho^4 \Xi_a^2 \Xi_b^2}, \\
& g_{03}=g_{30}=-\frac{\Delta_\theta a\left(2 m \rho^2\right) \sin ^2 \theta}{\rho^4 \Xi_a^2 \Xi_b}, \\
& g_{04}=g_{40}=-\frac{\Delta_\theta b\left(2 m \rho^2\right) \cos ^2 \theta}{\rho^4 \Xi_b^2 \Xi_a}, \\
& g_{11}=\frac{\rho^2}{\Delta_r}, \quad g_{22}=\frac{\rho^2}{\Delta_\theta}, \\
& g_{33}=\frac{\left(r^2+a^2\right) \sin ^2 \theta}{\Xi_a}+\frac{a^2\left(2 m \rho^2\right) \sin ^4 \theta}{\rho^4 \Xi_a^2}, \\
& g_{44}=\frac{\left(r^2+b^2\right) \cos ^2 \theta}{\Xi_b}+\frac{b^2\left(2 m \rho^2\right) \cos ^4 \theta}{\rho^4 \Xi_b^2}, \\
& g_{34}=g_{43}=\frac{a b\left(2 m \rho^2\right) \sin ^2 \theta \cos ^2 \theta}{\rho^4 \Xi_a \Xi_b} .
\end{aligned}
\ee
The temperature is given by:
\be
\label{eq:KerrAdS5_Temperature}
T=\frac{r_{+}^4\left[1+g^2\left(2 r_{+}^2+a^2+b^2\right)\right]-(a b)^2}{2 \pi r_{+}\left[\left(r_{+}^2+a^2\right)\left(r_{+}^2+b^2\right)\right]}.
\ee
For simplicity, we consider the $a=b$ case first in the rest parts of this section.

\subsection{Ensemble choice and near-horizon limit}
Under the extreme condition, we have
\bea
m_0=2 r_0^2 \left(1 + g^2 r_0^2\right)^3,\\
a_0=\sqrt{r_0^2 + 2 g^2 r_0^4},
\eea
where $r_0$ is the radius of the extreme black hole horizon. In our ensemble, we choose $a$ to be 
a constant. $m$ and $r_+$, the radius of the outer horizon, depend on $T$:
\bea
m(T)=m_0+c_1 T+c_2 T^2+c_3 T^3+O(T^4),\\
r_+(T)=r_0+d_1 T+d_2 T^2+d_3 T^3+O(T^4),
\eea
where the coefficients $c_i,d_i$ can be obtained by solving  \eqref{eq:KerrAdS5_Temperature} and $\Delta_{r}=0$. 
We can see $c_1=0$. Other coefficients are lengthy and not illuminating to be written here. The Hawking entropy can be calculated using the above temperature dependence:
\be
\label{eq:KerrAdS5_Classical Entropy}
S=\frac{A}{4}=\frac{2 \pi^2 r_0^3}{(1 - 2 g^2 r_0^2)^2} + \frac{4 \pi^3 r_0^4 (1 - g^2 r_0^2) }{1 - 12 g^4 r_0^4 + 16 g^6 r_0^6}T+O(T^2).
\ee
To zoom into the near-horizon region, we conduct the following coordinate transformation:
\be
r=r_+(T)+T k_1(y-1),\ t=\frac{-i\tau}{k_2 T},\ \phi_1=\varphi_1+\frac{-ik_3\tau}{T}+k_4i\tau,\ \phi_2=\varphi_2+\frac{-ik_5\tau}{T}+k_6i\tau,
\ee
where 
\be
\begin{aligned}
k_1 &= \frac{2 \pi r_0^2 (1 + g^2 r_0^2)}{1 + 4 g^2 r_0^2}, \\
k_2 &= 2 \pi, \\
k_3 &= \frac{\sqrt{r_0^2 + 2 g^2 r_0^4}}{4 \pi r_0^2}, \\
k_4 &= \frac{-\left(-1 + 2 g^2 r_0^2\right) \sqrt{r_0^2 + 2 g^2 r_0^4}}{2 \left(r_0 + 4 g^2 r_0^3\right)}, \\
k_5 &= k_3, \\
k_6 &= k_4.
\end{aligned}
\ee
By taking the $T\to0$ limit, we obtain the near-horizon geometry on the zeroth order of $T$:
\be
\label{eq:Kerr-AdS5 NearHorizon Zeroth}
\Bar{g}_{\mu\nu}dx^\mu dx^\nu =  g_1\left(\frac{dy^2}{y^2-1} + d\tau ^2 \left(y^2-1\right)\right)+   g_2d\theta^2 +g_{ij}( \theta) \Big(d x^i+i k_i(y-1)  d\tau   \Big) \Big(d x^j+i k_j(y-1)  d\tau   \Big),
\ee
where
\be
\begin{aligned}
g_1=&\frac{r_0^2}{2 \left(1 + 4 g^2 r_0^2\right)}\\
g_2=&\frac{2 r_0^2}{1 - 2 g^2 r_0^2}\\
g_{\varphi_1\varphi_1}(\theta)=&-\frac{r_0^2 \left(-3 + 2 g^2 r_0^2 + \left(1 + 2 g^2 r_0^2\right) \cos(2 \theta)\right) \sin^2(\theta)}{\left(1 - 2 g^2 r_0^2\right)^2}\\
g_{\varphi_1\varphi_2}(\theta)=&\frac{2 r_0^2 \left(1 + 2 g^2 r_0^2\right) \cos^2(\theta) \sin^2(\theta)}{\left(1 - 2 g^2 r_0^2\right)^2}\\
g_{\varphi_2\varphi_2}(\theta)=&\frac{r_0^2 \cos^2(\theta) \left(3 - 2 g^2 r_0^2 + \left(1 + 2 g^2 r_0^2\right) \cos(2 \theta)\right)}{\left(1 - 2 g^2 r_0^2\right)^2}
\\
k_{\varphi_1}=k_{\varphi_2}=&\frac{\left(-1 + 2 g^2 r_0^2\right) \sqrt{r_0^2 + 2 g^2 r_0^4}}{2 \left(r_0 + 4 g^2 r_0^3\right)}\\
\end{aligned}
\ee
By keeping the first order of $T$, we can also obtain $\delta g_{\mu\nu}$, the near-horizon geometry on $O(T)$. Again, this expression is lengthy and not illuminating to be written here.

\subsection{Lichnerowicz operator and the zero modes}
The Lichnerowicz operator of our Einstein-Hilbert action with cosmology constant is as follows:
\bea
\notag
h_{\alpha\beta}\Delta^{\alpha \beta, \mu \nu}_{L}h_{\mu\nu} &=& -\frac{1}{16\pi}h_{\alpha \beta}\Bigg(\frac{1}{2} g^{\alpha \mu} g^{\beta \nu} \square-\frac{1}{4} g^{\alpha \beta} g^{\mu \nu} \square+ R^{\alpha \mu \beta \nu}+ R^{\alpha \mu} g^{\beta \nu}-R^{\alpha \beta} g^{\mu \nu}-\frac{1}{2} R g^{\alpha \mu} g^{\beta \nu} \\
&+&\frac{1}{4} R g^{\alpha \beta} g^{\mu \nu}+\Lambda g^{\alpha \mu}g^{\beta\nu}-\frac{1}{2}\Lambda g^{\alpha \beta}g^{\mu\nu}\Bigg) h_{\mu \nu}.
\label{eq:KerrAdS5 Linch operator}
\eea
At zero temperature when $g=\bar{g}$, one can verify that the above operator has the following zero modes:
\be
h_{\mu\nu}^{(n)}dx^\mu dx^\nu=c_n e^{i n \tau} \left(\frac{-1 + y}{1 + y}\right)^{\frac{\abs{n}}{2}} (d\tau^2-2i\frac{\abs{n}}{n}\frac{d\tau dy}{y^2-1}-\frac{dy^2}{(y^2-1)^2})\quad \abs{n}\geq2,
\label{eq:KerrAdS5 zero modes}
\ee
where $c_n$ is the normalization constant.

\subsection{The lifted eigenvalue and entropy correction}
By turning on a small temperature, the "zero modes" will be lifted and gain a nonzero eigenvalue. Now we substitute $g=\bar{g}+\delta g$ into the Lichnerowicz operator \eqref{eq:KerrAdS5 Linch operator} and apply it to the zero modes \eqref{eq:KerrAdS5 zero modes}. The lifted eigenvalue is
\be
\delta\lambda_n=\int dx^4 \sqrt{\bar{g}}h^{(n)*}_{\alpha\beta}\delta\Delta_L^{\alpha\beta,\mu\nu} h^{(n)}_{\mu\nu}.
\ee
The expression of the operator is intractable, but the integration is straight forward and the result is simple
\be \label{eq:liftedeigenvalue_AdS5Kerr}
\delta\lambda_n=nT \frac{1-g^2r_0^2}{8r_0^2},\ \abs{n}\geq2.
\ee
The contribution of the extremal zero modes to the low-temperature partition function is therefore:
\be
\delta \log Z=\log(\prod_{n\geq2}\frac{\pi}{\lambda_n})=\frac{3}{2}\log (\frac{T}{T_q})+O(1).
\ee
Again, the $O(1)$ term is not reliable, and $T_q$ can be extracted from the $O(T)$ term of the classical entropy \eqref{eq:KerrAdS5_Classical Entropy}:
\be
T_q=\frac{1 - 12 g^4 r_0^4 + 16 g^6 r_0^6}{4 \pi^3 r_0^4 (1 - g^2 r_0^2)}.
\ee
%%%%%%%%%%%%%%%%%%%%%%%%%%%%%%%%%%%%%%%%%%
\subsection{Results for different angular momenta}
For completeness, we also do the calculation for different angular momenta. Central results are presented in this subsection. The Kerr-AdS black hole in five dimensions is naturally parameterized by $(a,b,m)$. $r_0$ and $m_0$ depend on $a,b$. However, for convenience, we use $(r_0,b,m)$ instead of $(a,b,m)$ as parameters in our calculation. The zeroth order near-horizon metric is given by:
\be
\label{eq:Kerr-AdS5_DiffAngularMomenta NearHorizon Zeroth}
\Bar{g}_{\mu\nu}dx^\mu dx^\nu =  g_1\left(\frac{dy^2}{y^2-1} + d\tau ^2 \left(y^2-1\right)\right)+   g_2d\theta^2 +g_{ij}( \theta) \Big(d x^i+i k_i(y-1)  d\tau   \Big) \Big(d x^j+i k_j(y-1)  d\tau   \Big),
\ee
where the coefficients are 
\be
\begin{aligned}
g_1=&-\frac{(b^2 + r_0^2) \left(-b^2 - r_0^2 + (b^2 - r_0^2 - 2 g^2 r_0^4) \cos(2 \theta)\right)}{8 \left(b^4 g^2 - g^4 r_0^6 + b^2 (1 + 3 g^2 r_0^2)\right)},\\
g_2=&\frac{(b^2 + r_0^2) \left(-b^2 - r_0^2 + (b^2 - r_0^2 - 2 g^2 r_0^4) \cos(2 \theta)\right)}{2 \left(g^2 r_0^4 (1 + b^2 g^2 + 2 g^2 r_0^2) \cos^2\theta + (b^2 - g^2 r_0^4) (-1 + b^2 g^2 \sin^2\theta)\right)},\\
g_{\varphi_1\varphi_2}(\theta)=&\frac{2 b r_0^2 (b^2 + r_0^2) (1 + g^2 r_0^2) \sqrt{1 + b^2 g^2 + 2 g^2 r_0^2} \sqrt{b^2 - g^2 r_0^4} \cos^2\theta \sin^2\theta}{(-1 + b^2 g^2) (2 g^2 r_0^4 + b^2 (-1 + g^2 r_0^2)) (b^2 + r_0^2 + (-b^2 + r_0^2 + 2 g^2 r_0^4) \cos(2 \theta))},\\
g_{\varphi_1\varphi_1}(\theta)=&-\bigg(r_0^2 (b^2 + r_0^2) (-b^4 + g^2 r_0^6 (3 + 2 g^2 r_0^2) + b^2 r_0^2 (-2 + g^2 r_0^2 + g^4 r_0^4) + (b^4 + g^2 r_0^6 \\
&+ 2 g^4 r_0^8 + b^2 g^2 r_0^4 (-1 + g^2 r_0^2)) \cos(2 \theta)) \sin^2\theta\bigg)/\bigg(\left(2 g^2 r_0^4 + b^2 (-1 + g^2 r_0^2)\right)^2\\ &\left(b^2 + r_0^2 + (-b^2 + r_0^2 + 2 g^2 r_0^4) \cos(2 \theta)\right)\bigg),\\
g_{\varphi_2\varphi_2}(\theta)=&\frac{2 b r_0^2 (b^2 + r_0^2) (1 + g^2 r_0^2) \sqrt{1 + b^2 g^2 + 2 g^2 r_0^2} \sqrt{b^2 - g^2 r_0^4} \cos^2\theta \sin^2\theta}{(-1 + b^2 g^2) (2 g^2 r_0^4 + b^2 (-1 + g^2 r_0^2)) (b^2 + r_0^2 + (-b^2 + r_0^2 + 2 g^2 r_0^4) \cos(2 \theta))},\\
k_{\varphi_1}=&-\frac{\sqrt{1 + b^2 g^2 + 2 g^2 r_0^2} \left(2 g^4 r_0^8 + b^4 (1 - g^2 r_0^2) + b^2 g^2 r_0^4 (-3 + g^2 r_0^2)\right)}{2 r_0 \sqrt{b^2 - g^2 r_0^4} \left(b^4 g^2 - g^4 r_0^6 + b^2 (1 + 3 g^2 r_0^2)\right)},\\
k_{\varphi_2}=&\frac{b (-1 + b^2 g^2) r_0 (1 + g^2 r_0^2)}{2 \left(b^4 g^2 - g^4 r_0^6 + b^2 (1 + 3 g^2 r_0^2)\right)}.
\end{aligned}
\ee
The metric \eqref{eq:Kerr-AdS5_DiffAngularMomenta NearHorizon Zeroth} is written in the generic form shown to admit a $SL(2,\mathbb{R})$ subgroup of isometries via a lemma in \cite{Kunduri:2007vf} which we quote explicitly in appendix \ref{App:SL2R}. 

The zero modes are given by
\bea
\notag
h_{\mu\nu}^{(n)}dx^\mu dx^\nu=&c_n& e^{i n \tau}\left(b^2 - r_0^2 + \left(b^2 - r_0^2 - 2 g^2 r_0^4\right) \cos(2 \theta)\right) \left(\frac{-1 + y}{1 + y}\right)^{\frac{\abs{n}}{2}}\\
&(d\tau^2&-2i\frac{\abs{n}}{n}\frac{d\tau dy}{y^2-1}-\frac{dy^2}{(y^2-1)^2}),\quad \abs{n}\geq2.
\label{eq:KerrAdS5_DiffAngularMomenta zero modes}
\eea
As final result we obtain
\be
\delta \log Z=\frac{3}{2}\log (\frac{T}{T_q})+O(1).
\ee
%%%%%%%%%%%%%%%%%%%%%%%%%%%%%%%%%%%%%%%%%%
\subsection{Comments on the rotating black hole in minimal gauged supergravity}

The rotating, electrically charged black hole in minimal gauged supergravity  in five dimensions plays an important role in the  context of the AdS/CFT correspondence. Importantly,  there is an explanation for its Bekenstein-Hawking entropy in terms of the superconformal index of ${\cal N}=4$ SYM \cite{Cabo-Bizet:2018ehj, Choi:2018hmj, Benini:2018ywd}. Some aspects of the logarithmic corrections to the entropy ($\log S_0$) have been discussed on the field theory side \cite{GonzalezLezcano:2020yeb}.Some of these logarithmic corrections to the entropy have been matched by using generalizations of the Kerr/CFT correspondence at one-loop level in   \cite{David:2021qaa}. We have performed some preliminary explorations  of the logarithmic in temperature corrections to the entropy in these models. The results are consistent with other models studied in this manuscript but will be reported elsewhere with a discussions of all the other zero modes arising from the gravitino and other supergravity fields.

%%%%%%%%%%%%%%%%%%%%%%%%%%%%%%%%%%%%%%%%%%%%%%%%%%
\section{On the origin of the universal $\frac{3}{2} \log T$ behavior}\label{Sec:Universality}

Having performed a number of explicit computations for various spacetimes in dimensions four and five, we are ready to highlight aspects of the universality of the logarithmic in temperature tensor modes contributions to the thermodynamics of generic near-extremal black holes.  The (mathematical) origin of $\frac{3}{2}$ factor of $\frac{3}{2}\log T_{\rm Hawking}$ behavior can be understood as a consequence of the following universal facts and general arguments.  They are elucidated sequentially as follows. 

\begin{enumerate}
    \item Extremal black holes in their near horizon geometry  always have an AdS$_2$ factor, possibly fibered over some compact directions. This is explicitly seen in \eqref{eq:zerothKerrAdSmetric} for the Kerr-AdS case and in \eqref{eq:KN-AdS4_NearHorizon Zeroth} for the Kerr-Newman AdS, similarly, it can be seen in the five-dimensional cases we considered \eqref{eq:NHEK Myers Perry} and \eqref{eq:Kerr-AdS5 NearHorizon Zeroth}. This  behavior is expected following general arguments given in \cite{Sen:2012cj,Bardeen:1999px,Astefanesei:2006dd,Kunduri:2007vf,Kunduri:2008rs,Figueras:2008qh}, in the particular case of Kerr black holes, the explicit form of the near-horizon isometry group was first explicitly presented in \cite{Bardeen:1999px}. This implies that there is a near horizon enhancement in the isometry group to $G=SL(2,\mathbb{R})\times G_{compact}$ where $G_{compact}$ depends on the black hole in question, whose near horizon we are interested in. In what follows, the exact nature of $G_{compact}$ will play no role and we focus on the fact that there exists an unbroken $SL(2,\mathbb{R})$ at the near horizon. 
    \item Technically, the $\log T$ corrections arise as corrections to the eigenspectrum of the Lichnerowicz operator. This follows from perturbation theory. Namely, imagine that we know the eigenspectrum $\{h_n, \lambda^0_n \}$ of the Lichnerowicz operator $\bar \Delta$ evaluated at the \textit{NHEK-AdS}. This gives
    \begin{eqnarray} \label{eq:eveqnLichOp}
        \bar \Delta\, h_n= \lambda^0_n h_n\, . 
    \end{eqnarray}
    Now we move slightly away from extremality, i.e., we turn on a small temperature $T$. This induces a change in the metric $\bar g \rightarrow g = \bar g + T\,\delta g + O(T^2)$. This in turn induces a change in the Lichnerowicz operator $\bar \Delta \rightarrow \Delta = \bar \Delta + \delta \Delta (T)$ and the eigenspectrum $\{h_n + \delta h_n  (T), \lambda^0_n + \delta \lambda_n  (T)\}$. First order perturbation theory then tells us that the corrected eigenvalue is the expectation value of the corrected operator. That is,
    \begin{eqnarray}
        \delta \lambda_n  (T) = \int d^4x \sqrt{\bar{g}}\: \left(h^{(n)}\right)_{\alpha\beta} \left(\delta \Delta\right)^{\alpha\beta, \mu\nu} \left(h^{(n)}\right)_{\mu\nu}\,.
    \end{eqnarray}
Since it is a Gaussian integral over bosonic fields, we have, ignoring an overall factor which play little role
    \begin{eqnarray} \label{eq:generalexpZ}
        Z \sim \prod_{n} \frac{1}{\left(\lambda^0_n + \delta \lambda_n  (T)\right)}\, , \qquad \log Z \sim - \sum_n \log \left(\lambda^0_n + \delta \lambda_n  (T) \right)\, , 
    \end{eqnarray} To proceed further, we need two crucial observations as argued in section \ref{Sec:Kerr-AdS}. Since  $\lambda_n^0$ is $T$-independent, $\log T$ corrections would necessarily come from the corrected eigenvalue $\delta\lambda_n(T)$. Also the Maclaurin expansion of $\log (a + x(T))$
    \begin{eqnarray}
        \log (a + x(T)) = \log a + \sum^\infty_{n=1} \frac{x^n(T)}{a^n n}\, . 
    \end{eqnarray} is a series in small $x\left(T\right)$, and therefore, the $\log x(T)$ corrections come only when $a=0$. Adapting this observation to our case, this implies that the $\log T $ corrections are obtained only from the zero modes of the Lichnerowicz operator when $\lambda^0_n=0$. In this situation, we have the following schematic form of the $\log T$ correction, given by
    \begin{eqnarray}
        \log Z \sim \log \prod_n \frac{1}{\delta \lambda_n(T)} 
    \end{eqnarray} where $\delta \lambda_n (T) = n f_1 T + O\left(T^2\right)$ . Note that, if $\delta \lambda_n (T)$ necessarily starts from $O(T)$ as any constant value would \textit{not} contribute to $\log T$ as per the discussion in this point. It is crucial to observe that the corrected eigenvalue scales linearly with $n$. This is seen in all our previous computations, for example, in \eqref{eq:correctedeigenval_KerrAdS} for the Kerr-AdS$_4$ black hole, in \eqref{eq:correctedeigenval_KerrAdSNewmanGaugedSUGRA} for the Kerr-Newman AdS black hole in $\mathcal{N}=4$ gauged supergravity, in \eqref{eq: LiftedEigenvalueMyersPerry_ II} for the Myers-Perry black hole in $5d$, and finally in \eqref{eq:liftedeigenvalue_AdS5Kerr} for the AdS$_5$ Kerr black hole. So far, the evidence has been empirical observation for all the sets of black holes that we have considered in this note. Later on, in this section, we shall motivate this further, but for the time being, let us accept that the corrected eigenvalues scale linearly with $n$.  

    \item We need a third ingredient to establish our claim that we have stated at the beginning of this section. This involves realising the  range for $n$ in equation \eqref{eq:generalexpZ} and this is where point (1) above comes in. As described there, the near horizon geometry necessarily has an AdS$_2$ factor. The eigenfunctions $h^0_n$ of the Lichnerowicz operator in equation \eqref{eq:eveqnLichOp} are generated by diffeomorphisms. Namely, 
    \begin{eqnarray}
        h_n \sim \mathcal{L}_{\xi^{(n)}} \bar g\, . 
    \end{eqnarray} Now, due to the presence of the AdS$_2$ throat, $\mathcal{L}_{\xi^{(n)}} \bar g$ vanish for $n=\pm 1, 0$ as they correspond to the isometry of AdS$_2$. Note that this fact, again, is universal due to the argument present in point (1). For example, this expectation has been borne out in \eqref{eq:zerothKerrAdSmetric} for the $4d$ Kerr-AdS black hole, in \eqref{eq:BPS-AdS4 NearHorizon Zeroth} for the $4d$ Kerr-Newman black hole in $\mathcal{N}=4$ gauged supergravity, in \eqref{eq:NHEK Myers Perry} for the $5d$ Myers-Perry black hole, in \eqref{eq:Kerr-AdS5 NearHorizon Zeroth} for the AdS$_5$ Kerr black hole.  The presence of an AdS$_2$ throat in the near horizon of all the black hole geometries we have studied implies $n$ runs from $|n|\geq 2$, as argued above. This immediately gives us 
    \begin{eqnarray}
        \log Z \sim \log \prod^\infty_{n=2} \frac{1}{n f_1 T}\,.
    \end{eqnarray} To regularize the above result , we use Zeta function regularization \cite{GonzalezLezcano:2023cuh} that gives
    \begin{eqnarray}
        \prod_{n=2}^{\infty} \frac{\alpha}{n} \sim \alpha^{-\frac{3}{2}} \,  . 
    \end{eqnarray}
    Setting $\alpha=\frac{1}{T}$, this immediately gives us the $\log T$ correction to be
    \begin{eqnarray}
        \log Z \sim \frac{3}{2} \log T + \ldots 
    \end{eqnarray}
 Note, from the arguments above, the presence of the factor $\frac{3}{2}$ in the $
 \log T$ correction relies just on two universal facts, which where shown to be manifestly true in all the cases we have analysed above, viz.,
 \begin{enumerate}
      \item Presence of an AdS$_2$ throat in the near horizon geoemtry, and
      \item Regularized value of an infinite product.
  \end{enumerate} Since the above two points are independent of the black hole geometry in question, the presence of an universal $\frac{3}{2}$ factor in the $\log T$ correction is strongly motivated. 
\end{enumerate}
\paragraph{Insensitive to matter sector:} One could wonder if the presence of matter affects the argument presented above. Certainly, it is true, that in presence of matter, there could be some other contribution over and above the one discussed above. However, we can go beyond and motivate that the $\log T$ receives no correction from the matter sector.  This follows from observing the terms that actually contribute to the corrected eigenvalue of the zero mode  \eqref{eqn: LiftedEigenvalue}. The contributions come from only two terms in $h_{\alpha\beta} \delta \Delta^{\alpha\beta, \mu\nu} h_{\mu\nu}$, viz., from 
\begin{itemize}
    \item $ h_{\alpha\beta}\ \delta\left(\frac{1}{2}g^{\alpha\mu}g^{\beta\nu}\square h_{\mu\nu}\right) $
    \item $h_{\alpha\beta}\ \delta \left(R^{\alpha\mu\beta\nu}\right)h_{\mu\nu}$
\end{itemize} For example, this claim can be checked from \eqref{eq:cancelling terms KN-AdS4} and we commented on this fact below \eqref{eq:Cancellation}. This implies that the contribution comes only from the metric data and is independent of the matter sector and the arguments presented above lead us to conclude that the $\frac{3}{2} \log T$ correction is universal and linked solely to the presence of near horizon AdS$_2$ throat. 
%%%%%%%%%%%%%%%%%%%%%%%%%%%%%%%%%%%%%%%%%%%%%%%%%%%%%%
\section{Conclusions}\label{Sec:Conclusions}

In this manuscript we have systematically discussed the low-temperature quantum contributions to the partition functions of near-extremal rotating black holes. In particular, we have explicitly treated various black holes in asymptotically four-dimensional AdS spacetime: the Kerr-AdS$_4$ black hole in section \ref{Sec:Kerr-AdS}, the Kerr-Newman-AdS$_4$ black hole in section \ref{Sec:KN-AdS} and the rotating black hole with two electric charges in ${\cal N}=4$ gauged supergravity in section \ref{Sec:Sugra}. In five dimensions we have addressed the asymptotically flat Myers-Perry black hole  in section \ref{Sec:MyersPerryBlack hole} and the Kerr-AdS$_5$ in section \ref{sec:Kerr-AdS5}. In every case we found that {\it the contribution to the path integral arising from the gravitational tensor modes is  universal and equal to} $\frac{3}{2}\log T_{\rm Hawking}$. 

With the above overwhelming amount of evidence at hand, we argue for this universality based on two main facts: (i) The near-horizon geometry contains the $SL(2,\mathbb{R})$ group of isometries and (ii) The relevant contribution from the Lichnerowicz operator is matter-independent;  the answer, in all cases, comes from only two geometric terms: $g^{\alpha \mu}g^{\beta\nu}\square$ and $R^{\alpha \mu\beta\nu}$. Our computations explicitly demonstrate a direct and universal embedding of an important quantum aspect, first discussed in the context of JT gravity, into higher-dimensional rotating black holes in both asymptotically flat and asymptotically AdS spacetimes.

There is a number of interesting problems that our result motivates. It would be interesting to track our arguments in a more rigorous fashion. Black hole dynamics is highly dependent on the dimensionality of spacetime, the universality discussed in this paper, being quantum, opens the door for other similar universal properties for all near-extremal black holes in various dimensions.

Given recent advances in the understanding of the microscopic entropy of asymptotically AdS black holes, it would be interesting to consider their partition functions in the near-extremal context along the lines pursued in this manuscript.  In section \ref{Sec:Sugra} we have discussed only the tensor modes for the two-charge, rotating black hole in ${\cal N}=4$ gauged supergravity whose field theory dual is well understood; a complete analysis will require a proper treatment of the vector and fermionic zero modes. One expects that in these supersymmetric cases the relevant intuition should arise from supersymmetric versions of JT gravity. Indeed, some relevant work has already been presented in \cite{Boruch:2022tno} and it would be interesting to extend it to other rotating black holes and to more general configurations. 

Finally, it remains a truly interesting problem to derive this low-temperature thermodynamics directly from field theory in the context of the AdS/CFT correspondence.  A natural starting point is to consider the reduction of the dual field theory to one-dimension. For example, in \cite{Benini:2022bwa}, the reduction of the three-dimensional theory dual to certain magnetically charged asymptotically AdS$_4$ black hole  was performed but extracting the expected behavior might prove a Herculean task. It is also possible to consider bottom-up toy models arising from modifications of the Sachdev-Ye-Kitaev models which contain the Schwarzian mode that figures so prominently on the gravity side. This direction holds the promise to address some of the puzzles (including the factorization puzzle) that low-dimensional models of gravity such as JT gravity, have  brought upon us.

%%%%%%%%%%%%%%%%%%%%%%%%%%%%%%%%%%%%%%%%%%%%%%%%%%%%%%
\section*{Acknowledgments}
We are grateful to Alejandra Castro, Marina David, Matt Heydeman, Imtak Jeon, Alfredo Gonz\`alez Lezcano,  Juan Maldacena, Jun Nian and  Joaqu\'\i n Turiaci for discussions on related topics. This work is supported in part by the U.S. Department of Energy under grant DE-SC0007859. LPZ gratefully acknowledges support from an IBM Einstein fellowship during 2022/2023 while at the Institute for Advanced Study. AR is supported by an appointment to the JRG program at the APCTP through the Science and Technology Promotion Fund and the Lottery Fund of the Korean Government, by the Korean Local Governments - Gyeongsangbuk-do Province and Pohang City and by the National Research Foundation of Korea (NRF) grant funded by the Korean government (MSIT) (No. 2021R1F1A1048531). SM acknowledges the warm hospitality of APCTP, Postech, South Korea during the APCTP Winter School on Fundamental Physics 2024, where a part of this work was carried out. SM is financially supported by the institute post-doctoral fellowship of IITK.

\appendix

%%%%%%%%%%%%%%%%%%%%%%%%%%%%%%%%%%%%%%%%%%%%%%%%%%%%%%%%%
 \section{Kerr-AdS$_{4}$: Grand canonical ensemble}\label{appendix:againKerrAdS4}
 In this appendix, we provide the parallel story for Kerr-AdS black hole in a different, grand canonical ensemble choice from that in section \ref{Sec:Kerr-AdS}. We start with the Einstein-Hilbert action:
\be
 I = -\frac{1}{16\pi} \int_{\mathcal{M}} d^4x \sqrt{-g} \left(R - 2\Lambda \right)+I_{boundary}.
\ee
The Kerr-Newman-AdS$_4$ solution is given by 
\begin{equation} \label{eq:metricKerrAdS4_another ensemble}
    ds^2 = - \frac{\Delta_r}{\Sigma^2} \left(dt - \frac{a}{\Xi} \sin^2 \theta d\phi\right)^2 + \frac{\Sigma^2}{\Delta_r}dr^2 + \frac{\Sigma^2}{\Delta_\theta}d\theta^2 + \frac{\Delta_\theta}{\Sigma^2}\sin^2 \theta\left( a dt- \frac{\left(r^2+a^2\right)}{\Xi}d\phi \right)^2 \, ,
\end{equation}
where
\begin{eqnarray}
    \Delta_r &=& \left(r^2+a^2\right)\left(a+g^2r^2\right) - 2 M r \, , \quad  \Delta_\theta = 1 - g^2a^2 \cos^2\theta\, , \\
    \Xi &=& 1 - g^2a^2 \, ,  \qquad \Sigma = \sqrt{r^2 + a^2 \cos^2 \theta}\, . 
\end{eqnarray}
The cosmological constant $\Lambda$ is related $g\equiv\frac{1}{L}$ by $\Lambda = -3g^2$.
%%%%%%%%%%%%%%%%%%%%%%%%%%%%%%%%%%%%%%%%%%%
\subsection{Ensemble choice and near-horizon limit}
In this section, we illustrate our ensemble choice when turning on a small temperature as we zoom into the near-horizon region of the Kerr-AdS geometry. We obtain the near-horizon geometry on the zeroth and the first orders of temperature.

At extremality  we have 
\bea
m_0=-\frac{r_0 (1 + g^2 r_0^2)^2}{-1 + g^2 r_0^2},\\
a_0=g r_0 \sqrt{\frac{1 + 3 g^2 r_0^2}{g^2 - g^4 r_0^2}},
\eea
where $r_0$ is the radius of the extreme black hole horizon, $g=\frac{1}{L}$ and $L$ is the AdS radius. In our ensemble, we choose $m$ to be constant, $a$ and $r_+$, the radius of the outer horizon, depend on $T$,
\bea
a(T)=a_0+c_1 T+c_2 T^2+c_3 T^3+O(T^4),\\
r_+(T)=r_0+d_1 T+d_2 T^2+d_3 T^3+O(T^4),
\eea
Where the coefficients $c_i,d_i$ can be obtained by solving the second equation of \eqref{eq:Temperature of BPS} and $\Delta_{r}=0$. 
We verify that $c_1=0$. Other coefficients are lengthy and not illuminating to be written here. The Hawking entropy can be calculated using the above temperature dependence,
\be
\label{eq:KerrAdS_Classical Entropy}
S=\frac{2 \pi r_0^2}{1 - 3 g^2 r_0^2}+\frac{8 \pi^2 r_0^3 (1 - g^2 r_0^2)}{1 + 3 g^2 r_0^2 - 21 g^4 r_0^4 + 9 g^6 r_0^6}T+O(T^2).
\ee
To zoom into the near-horizon region, we conduct the following coordinate transformation:
\be
r=r_+(T)+T k_1(y-1),\ t=\frac{-i\tau}{k_2 T},\ \phi=\varphi+\frac{-ik_3\tau}{T}+k_4i\tau,
\ee
where
\begin{align}
k_1 &= \frac{4 \pi r_0^2 (1 + g^2 r_0^2)}{1 + 6 g^2 r_0^2 - 3 g^4 r_0^4}, \\
k_2 &= 2 \pi, \\
k_3 &= -\frac{g \sqrt{\frac{-1 - 3 g^2 r_0^2}{g^2 (-1 + g^2 r_0^2)}} (-1 + 3 g^2 r_0^2)}{2 k_2 r_0}, \\
k_4 &= -\frac{g (-1 + g^2 r_0^2) (-1 + 3 g^2 r_0^2) \sqrt{\frac{1 + 3 g^2 r_0^2}{g^2 - g^4 r_0^2}}}{-1 - 6 g^2 r_0^2 + 3 g^4 r_0^4}.
\end{align}
By taking the $T\to0$ limit, we obtain the near-horizon geometry on the zeroth order of $T$:
\be
\Bar{g}_{\mu\nu}dx^\mu dx^\nu =  g_1(\theta ) \left(\frac{dy^2}{y^2-1} + d\tau ^2 \left(y^2-1\right)\right)+   g_2( \theta)d\theta^2 +g_3( \theta) \Big(d \varphi+i k(y-1)  d\tau   \Big)^2,
\ee
where
\be
\begin{aligned}
    g_1(\theta)=&- \frac{r_0^2 \left(3 + g^2 r_0^2 + (1 + 3 g^2 r_0^2) \cos(2 \theta) \right)}{-2 - 12 g^2 r_0^2 + 6 g^4 r_0^4}, \\
    g_2(\theta)=&- \frac{r_0^2 \left(3 + g^2 r_0^2 + (1 + 3 g^2 r_0^2) \cos(2 \theta)\right)}{2 \left(-1 + g^2 r_0^2 + g^2 r_0^2 (1 + 3 g^2 r_0^2) \cos^2\theta\right)},\\
    g_3(\theta)=&-\frac{2 r_0^2 \left(-2 + 3 g^2 r_0^2 + 3 g^4 r_0^4 + g^2 r_0^2 \left(1 + 3 g^2 r_0^2\right) \cos(2 \theta)\right) \sin^2\theta}{\left(1 - 3 g^2 r_0^2\right)^2 \left(1 - g^2 r_0^2 + \left(1 + 3 g^2 r_0^2\right) \cos^2\theta\right)},\\
    k=&-k_4.
\end{aligned}
\ee
By keeping the first order in $T$, we obtain $\delta g_{\mu\nu}$, the near-horizon geometry on $O(T)$. This expression is lengthy and not illuminating to be written here.
\subsection{Lichnerowicz operator and the zero modes}
The Lichnerowicz operator of our Einstein-Hilbert action with cosmology constant is as follows:
\bea
\notag
h_{\alpha\beta}\Delta^{\alpha \beta, \mu \nu}_{L}h_{\mu\nu} &=&-\frac{1}{16\pi} h_{\alpha \beta}(\frac{1}{2} g^{\alpha \mu} g^{\beta \nu} \square-\frac{1}{4} g^{\alpha \beta} g^{\mu \nu} \square+ R^{\alpha \mu \beta \nu}+ R^{\alpha \mu} g^{\beta \nu}-R^{\alpha \beta} g^{\mu \nu}-\frac{1}{2} R g^{\alpha \mu} g^{\beta \nu} \\
&+&\frac{1}{4} R g^{\alpha \beta} g^{\mu \nu}+\Lambda g^{\alpha \mu}g^{\beta\nu}-\frac{1}{2}\Lambda g^{\alpha \beta}g^{\mu\nu}) h_{\mu \nu}.
\label{eq:kerr-AdS Linch operator}
\eea
At zero temperature when $g=\bar{g}$, one can verify that the above operator has the following zero modes:
\bea
\begin{aligned}
h_{\mu\nu}^{(n)}dx^\mu dx^\nu=&c_n e^{i n \tau} \left(\frac{y-1}{y+1}\right)^{\frac{\abs{n}}{2}} \left(3 + g^2 r_0^2 + \left(1 + 3 g^2 r_0^2\right) \cos 2\theta \right) \\
&\left(d\tau^2-2i\frac{\abs{n}}{n}\frac{d\tau dy}{y^2-1}-\frac{dy^2}{(y^2-1)^2}\right),\quad \abs{n}\geq2,
\end{aligned}
\label{eq:Kerr-AdS zero modes}
\eea
where $c_n$ is the normalization constant.
\subsection{The lifted eigenvalue and entropy correction}
By turning on a small temperature, the ``zero modes" will be lifted and gain a nonzero eigenvalue. Now we substitute $g=\bar{g}+\delta g$ into the Lichnerowicz operator \eqref{eq:kerr-AdS Linch operator} and apply it to the zero modes \eqref{eq:Kerr-AdS zero modes}. The lifted eigenvalue is
\be
\delta\lambda_n=\int dx^4 \sqrt{\bar{g}}h^{(n)*}_{\alpha\beta}\delta\Delta_L^{\alpha\beta,\mu\nu} h^{(n)}_{\mu\nu}.
\ee
The expression of the operator is intractable, but the integration is straight forward and the result is simple,
\be
\delta\lambda_n=\frac{3 n (1 - g^2 r_0^2)T}{32 r_0},\ \abs{n}\geq2.
\ee
The contribution of the extremal zero modes to the low-temperature partition function is therefore:
\be
\delta \log Z=\log(\prod_{n\geq2}\frac{\pi}{\lambda_n})=\log (\prod_{n\geq2}\frac{32\pi r_0}{3n(1-g^2r_0^2)T}).
\ee
Using Zeta function regularization
\be
\prod_{n\geq2}\frac{\alpha}{n}=\frac{1}{\sqrt{2\pi}}\frac{1}{\alpha^{3/2}},
\ee
the one-loop correction is
\be
\delta \log Z=\frac{3}{2}\log (\frac{T}{T_q})+O(1),
\ee
where the $O(1)$ term is not reliable. The temperature scale $T_q$ is the temperature under which the above one-loop correction needs to be considered. $T_q$ can be extracted from the classical entropy \eqref{eq:KerrAdS_Classical Entropy}. As indicated in \cite{Rakic:2023vhv},
\be
T_q=\frac{1 + 3 g^2 r_0^2 - 21 g^4 r_0^4 + 9 g^6 r_0^6}{8 \pi^2 r_0^3 (1 - g^2 r_0^2)}.
\ee

%%%%%%%%%%%%%%%%%%%%%%%%%%%%%%%%%%%%%%%
\section{$SL(2,\mathbb{R})$ symmetries of the near-horizon geometries}\label{App:SL2R}

Let us see the action of $SL(2,\mathbb{R})$ explicitly. We recall \cite{Bardeen:1999px} showed that the near-horizon extremal Kerr metric

\bea
\label{Eq:NHEK-BH}
ds^2 &=& \left(\frac{1+\cos^2\theta}{2}\right)
\bigg(-\frac{r^2}{r_0^2}dt^2 +\frac{r_0^2}{r^2}dr^2 + d\theta^2 \bigg) +
\frac{2r_0^2\sin^2\theta}{1+\cos^2\theta}\left(d\phi + \frac{r}{r_0^2}dt\right)^2\,, \quad 
\eea
admits the following Killing vectors. The precise claim is that first, the metric \eqref{Eq:NHEK-BH} has enhanced symmetry with respect to the full Kerr metric. Namely, in addition to $\partial_t$ and $\partial_\phi$ of the original Kerr, it is invariant under $r\to \alpha\, r, \quad t\to t/\alpha$, for any constant $\alpha$. To see the full $SL(2,\mathbb{R})\times U(1)$ it is convenient to transform to global coordinate where $SL(2,\mathbb{R})$ arises as the isometries of AdS$_2$ and $U(1)$ is generated by $\partial_\phi$.

Every near-horizon metric we arrive at in this manuscript has a near horizon that is of the form $A.1$ and thus admit the action of $SL(2,\mathbb{R})\times U(1)$ as isometries.

Further improvement on this observation were presented in, for example, \cite{Kunduri:2007vf} that established that similar results more broadly for large classes of near-horizon backgrounds in fairly general theories, including in higher dimensions.

%%%%%%%%%%%%%%%%%%%%%%%%%%%%%%%%%%%%%%%%%%%%%
\subsection{Near-horizon symmetry of Kerr-AdS$_4$ black hole}
Here, we establish  the fact that our near horizon metric, as given in equation \eqref{eq:zerothKerrAdSmetric} is diffeomorphic to the metric given in equation (2) of \cite{Kunduri:2008rs}. Once established, this guarantees the existence of an isometry group $G=SL(2,\mathbb{R})\times U(1)$, as proven in \cite{Kunduri:2008rs}. The starting point for us is the \textit{NHEK-AdS} given in \eqref{eq:zerothKerrAdSmetric}. The computation proceeds in two steps. First we do a coordinate transformation such that the \textit{NHEK-AdS} metric may be written in the following manner
\begin{eqnarray}
  ds^2_{\textrm{\textit{NHEK-AdS}}} =  g_1(\theta ) \left(\frac{dy^2}{y^2-1} + d\tau ^2 \left(y^2-1\right)\right)+   g_3( \theta)^{-1}\text{d$ \tilde \theta $}^2 +g_3( \theta) \Big(d \phi+i g_4(y)(y-1)  d\tau   \Big)^2  \nonumber \\
\end{eqnarray} for some coordinate $\tilde \theta$  obtained as $ \theta \rightarrow \tilde \theta (\theta)$. This transformation is required to match the form given in \cite{Kunduri:2008rs}. To obtain the functional form of $\tilde \theta$, we introduce a coordinate transformation as  
\begin{eqnarray}
    \theta \rightarrow \tilde \theta (\theta): \, \quad g_2 (\theta) d\theta^2 = g_2 (\theta)  \tilde \theta ^\prime (\theta)^2 d\tilde\theta^2   
\end{eqnarray} and demand
\begin{eqnarray}
 g_2 (\theta)  \tilde \theta ^\prime (\theta)^2 =  g_3( \theta)^{-1}\, . 
\end{eqnarray} The above equation is solved to give 
\begin{eqnarray}
    \tilde \theta (\theta) = \int d{\theta} ~ \frac{1}{\sqrt{g_2 (\theta) g_3 (\theta)}} = - \frac{1-\frac{3r_0^2}{L^2} }{2r_0^2} \log(\cot{\frac{\theta}{2}})\, . 
\end{eqnarray}

For the second step, we see that the metric now has the general form which we write succinctly as
\begin{eqnarray} \label{eq:KerrNHEKgenform}
ds^2 = G_{11}  d\tau^2 + 
 G_{22}  dy^2 + G_{33} d\tilde \theta^2 + G_{44} d\phi^2 + 2 G_{14} d\tau d\phi 
\end{eqnarray} where, for example, $G_{33} = g_3^{-1} \left(\theta\right)$ and so on. To establish the desired diffeomorphism that takes the metric in equation \eqref{eq:KerrNHEKgenform} to the one in equation (2) of Lucietti et al, we do a further coordinate transformation given by
\begin{eqnarray}
    d\tau = dv + a dy\, , \qquad d\phi = d\varphi +b dy \, , 
\end{eqnarray}crucially without involving $\tilde \theta$. This induces the following change in the metric given by
\begin{eqnarray}
    &&ds^2 = G_{11} dv^2 + 2 \left(a G_{11} + b  G_{14}\right) dv dy + 2 G_{14} dv d\varphi + G_{33} d\theta^2 + G_{44} d\varphi^2 +\nonumber \\ 
    &&\big( G_{11} a^2 +  G_{44} b^2 + 
     2 G_{14} a b + G_{22} \big) dy^2 + 2(G_{14} a+ G_{44} b) d\varphi dy \, . \nonumber \\ 
\end{eqnarray} The next step is to choose $a$ and $b$ judiciously such that 
\begin{eqnarray}
(g_{\varphi y}:) &&\qquad    G_{14} a+ G_{44} b = 0 \, , \qquad \\
(g_{y y}:) &&\qquad      G_{11} a^2 +  G_{44} b^2 + 
     2 G_{14} a b + G_{22} = 0\, . 
\end{eqnarray}. The above pair of equations can simultaneously be solved for
\begin{eqnarray}
    a = \mp \frac{\sqrt{G_{22}}G_{44}}{G_{14} \sqrt{G_{44} - \frac{G_{11}G^2_{44} }{G^2_{14}}}}\, , \qquad   b =  \pm \frac{\sqrt{G_{22}}}{\sqrt{G_{44} - \frac{G_{11}G^2_{44} }{G^2_{14}}}}\, . 
\end{eqnarray} This choice of $a$ and $b$ takes our metric to the form stated in \cite{Kunduri:2008rs} and guarantees the existence of the isometry group $SL(2,\mathbb{R}) \times U(1)$.

%%%%%%%%%%%%%%%%%%%%%%%%%%%%%%%%%%%%%%%%%%%%%
\subsection{Near-horizon symmetry of Kerr-AdS black hole in various dimensions}

The situation in dimensions five, six and seven can be addressed with the help of the following 
{\bf Lemma \cite{Kunduri:2007vf}}: The metric

\be
ds^2 =\Gamma(\rho)\bigg[A_0 r^2 dv^2 +2dv dr\bigg] +d\rho^2 +\gamma_{ij}(\rho) (dx^i+k^i r dv)(dx^j+k^j r dv)
\ee
has isometry group $\hat{G}_3\times U(1)^{D-3}$ where the three-dimensional group $\hat{G}_3$ is 2d Poincar\`e if $A_0=0$ or $O(2,1)$ if $A_0\neq 0$. The orbits of $\hat{G}_3$ are three-dimensional if $k^i\neq 0$ and two-dimensional if $k^i=0$.

This powerful statement is central in arguing for the universality of the $\frac{3}{2}\log T$ contribution coming from the tensor modes. Coupling the existence of the $SL(2, \mathbb{R})$ subgroup of isometries with the fact that the contribution to the Lichnerowicz operator is strictly metric-driven, as the contributions from other matter fields cancel among themselves, we obtain necessary ingredients towards the universality of the $\frac{3}{2}\log T$ contribution in various dimensions.

%%%%%%%%%%%%%%%%%%%%%%%%%%%%%%%%%%%%%%%%%%%%%%%
\section{The structure of the Lichnerowicz operator}
\label{appendix:Lich operator}

The generalized Lichnerowicz operator is obtained by the second-order metric perturbation of the action. In this subsection, we list the ingredients of the operator used in the previous sections.

For $\mathcal{L}=R*1$, we can see Einstein-Hilbert's contribution to the Lichnerowicz operator is
\bea
\notag
h_{\alpha\beta}\Delta^{\alpha \beta, \mu \nu}_{EH}h_{\mu\nu}&=& h_{\alpha \beta}(\frac{1}{2} \bar{g}^{\alpha \mu} \bar{g}^{\beta \nu} \bar{\square}-\frac{1}{4} \bar{g}^{\alpha \beta} \bar{g}^{\mu \nu} \bar{\square}+ \bar{R}^{\alpha \mu \beta \nu}\\
&+& \bar{R}^{\alpha \mu} \bar{g}^{\beta \nu}-\bar{R}^{\alpha \beta} \bar{g}^{\mu \nu}-\frac{1}{2} \bar{R} \bar{g}^{\alpha \mu} \bar{g}^{\beta \nu}+\frac{1}{4} \bar{R} \bar{g}^{\alpha \beta} \bar{g}^{\mu \nu}) h_{\mu \nu}.
\eea
% \bea
% \frac{1}{2}g^{\alpha\mu}g^{\beta\nu}\square & =& \nonumber \\
% -\frac{1}{4}g^{\alpha\beta}g^{\mu\nu}\square & =& \nonumber \\
% R^{\alpha\mu\beta\nu}&=& \nonumber \\
% R^{\alpha\mu}g^{\beta\nu}&=& \nonumber \\
% -R^{\alpha\beta}g^{\mu\nu} &=& \nonumber \\
% -\frac{1}{2}R g^{\alpha \mu}g^{\beta\nu} &=& \nonumber \\
% \frac{1}{4}R g^{\alpha\beta}g^{\mu\nu}&=& \nonumber \\
% \eea

% For the simpler case of Kerr-AdS, the contributions are as follows:
For $\mathcal{L}=*F\wedge F=\frac{1}{2}*F^2$, the contribution is
\bea
\notag
h_{\alpha \beta} \Delta_{F}^{\alpha \beta, \mu \nu} h_{\mu \nu}&=&h_{\alpha \beta}(-\frac{1}{8} F^2\left(2 g^{\alpha \mu} g^{\beta \nu}-g^{\alpha \beta} g^{\mu \nu}\right)+ F^{\alpha \mu} F^{\beta \nu}\\
&+&2 F^{\alpha \gamma} F_{\ \gamma}^\mu g^{\beta \nu}-F^{\alpha \gamma} F_{\ \gamma}^\beta g^{\mu \nu}) h_{\mu \nu},
\eea
which can be applied to $F_1$ and $F_2$ in section \ref{Sec:Sugra}.

For $\mathcal{L}=*1$, the kinetic term is
\begin{equation}
    h_{\alpha \beta} \Delta_{1}^{\alpha \beta, \mu \nu} h_{\mu \nu}=h_{\alpha\beta}(-\frac{1}{2}g^{\alpha\mu}g^{\beta\nu}+\frac{1}{4}g^{\alpha\beta}g^{\mu\nu})h_{\mu\nu},
\end{equation}
which can be used to construct the contribution of cosmology constant.

For $\mathcal{L}=*d\zeta\wedge d\zeta$, the kinetic term is
\begin{equation}
    h_{\alpha \beta} \Delta_{\zeta}^{\alpha \beta, \mu \nu} h_{\mu \nu}=h_{\alpha\beta}(\nabla_\gamma\zeta\nabla^\gamma\zeta(-\frac{1}{2}g^{\alpha\mu}g^{\beta\nu}+\frac{1}{4}g^{\alpha\beta}g^{\mu\nu})+2g^{\alpha\mu}\nabla^\beta\zeta\nabla^\nu\zeta-g^{\alpha\beta}\nabla^\mu\zeta\nabla^\nu\zeta)h_{\mu\nu},
\end{equation}
which can also be applied to $\chi$ by $\zeta \leftrightarrow \chi$.

\bibliographystyle{JHEP}
\bibliography{LowT-AdS}
\end{document}